\documentclass[5p]{elsarticle}

\usepackage{amsmath}
\usepackage{graphicx}
\usepackage{color}
\usepackage{natbib}
\biboptions{square,numbers,sort&compress}

\newcommand{\vr}{\vv{r}}
\newcommand{\piv}{\Pi_\text{v}}
\newcommand{\pith}{\Pi_\text{th}}

\newcommand{\dsurd}[2]{\frac{\partial #1}{\partial #2}}
\newcommand{\sdsurd}[2]{\partial #1/\partial #2}

\newcommand{\vnabla}{\boldsymbol{\nabla}}
\newcommand{\vv}[1]{\mathbf{#1}}
\newcommand{\undemi}{\frac{1}{2}}

\newcommand{\conj}[1]{\overline{#1}}

\newcommand{\phivp}{\gamma}
\newcommand{\phipgradp}{\theta}
\newcommand{\ri}{x_i}
\newcommand{\pgradt}{\dsurd{\pact}{\ri}}
\newcommand{\pgrad}{G_i}

\newcommand{\vvr}{\vv{r}}
\newcommand{\pact}{p}
\newcommand{\pac}{P}
\newcommand{\firstpaper}{$[$OL~I$]$}
\newcommand{\tshift}{\tau^*}
\newcommand{\fbun}{\vv{F_{B}}}
\newcommand{\sstreamers}{S-streamers}
\newcommand{\lstreamers}{L-streamers}

\newcommand{\olchange}[1]{#1}

\begin{document}
%
\begin{frontmatter}
  \title{A simple model of ultrasound propagation in a cavitating
    liquid. Part II: Primary Bjerknes force and bubble structures.}
  
  
  \author[EMAC]{O. Louisnard\corref{CORRESPONDANT}}
  \ead{louisnar@enstimac.fr}
  \cortext[CORRESPONDANT]{Corresponding author}
  \address[EMAC]%
  {Centre RAPSODEE, FRE CNRS 3213, Universit\'e de Toulouse, %
    Ecole des Mines d'Albi, %
    \\  81013 Albi Cedex 09, France}
  

  \begin{abstract}
    In a companion paper, a reduced model for propagation of acoustic
    waves in a cloud of inertial cavitation bubbles was proposed. The
    wave attenuation was calculated directly from the energy
    dissipated by a single bubble, the latter being estimated directly
    from the fully nonlinear radial dynamics. The use of this model in
    a mono-dimensional configuration has shown that the attenuation
    near the vibrating emitter was much higher than predictions
    obtained from linear theory, and that this strong attenuation
    creates a large traveling wave contribution, even for closed
    domain where standing waves are normally expected. In this paper,
    we show that, owing to the appearance of traveling waves, the
    primary Bjerknes force near the emitter becomes very large and
    tends to expel the bubbles up to a stagnation point.
    Two-dimensional axi-symmetric computations of the acoustic field
    created by a large area immersed sonotrode are also performed, and
    the paths of the bubbles in the resulting Bjerknes force field are
    sketched. Cone bubble structures are recovered and compare
    reasonably well to reported experimental results. The underlying
    mechanisms yielding such structures is examined, and it is found
    that the conical structure is generic and results from the
    appearance a sound velocity gradient along the transducer area.
    Finally, a more complex system, similar to an ultrasonic bath, in
    which the sound field results from the flexural vibrations of a
    thin plate, is also simulated. The calculated bubble paths reveal
    the appearance of other commonly observed structures in such
    configurations, such as streamers and flare structures.
  \end{abstract}
  
\end{frontmatter}

\section{Introduction}
A common observation in acoustic cavitation experiments is the rapid
translational motion of the bubbles relative to the liquid, and their
self-organization into various spectacular structures. These
structures have been systematically reviewed recently
\cite{mettin2005}, and some of them have been successfully explained
by results derived from single bubble physics
\cite{parlitz99,mettin2005,mettin2007}.

The origin for bubble translational motion in an acoustic field is the
so-called Bjerknes force~\cite{bjerknes,zwick1,akhatov97une}, which is
the average over one oscillation period of the generalized buoyancy
force exerting on any body in an accelerating
liquid~\cite{magnaudeteames2000}. It is commonly expressed in terms of
the pressure gradient as:
\begin{equation}
  \label{bjerknes}
  \fbun = - \left<V \vnabla p  \right>,
\end{equation}
where $\left<. \right>$ denotes the average over one acoustic period,
$V$~is the bubble volume and $p$ the acoustic pressure which would
exist in the liquid at the center of the bubble if the latter were not
present. Since $V$ and $p$ are oscillatory, the average of their
product can be non-zero.  

The most spectacular and known manifestation of the primary Bjerknes
force occurs in standing waves. It can be simply deduced from linear
theory that bubbles smaller than the resonant size are attracted by
pressure antinodes, whereas bubbles larger than resonant size are
attracted by pressure nodes \cite{leightonwalton}. This was confirmed
by the early experiments of Crum \& Eller \cite{crumeller70}, and is
the basic principle of levitation experiments used to study single
bubble sonoluminescence, where attraction by the central antinode of
the flask counteracts the buoyancy force \cite{gaitancrum92, barber}.

However, nonlinear effects can produce repulsion of inertial bubbles
from pressure antinodes above a given threshold \cite{akhatov97une}.
This threshold can be estimated analytically for low frequency driving
and is found to be near 170~kPa, with a slight dependence on surface
tension \cite{louisnard2008}. This can be evidenced in multi-bubble
experiments by a void region near the pressure antinode surrounded by
bubbles accumulating near the threshold zone \cite{parlitz99}, or by
bubbles self-arrangement into parallel layers shifted relative to the
antinodal planes \cite{mettin2005}\olchange{, correctly predicted by
  particle model simulations.}

A more important issue concerns the Bjerknes force exerted on bubbles
by large amplitude traveling waves. While small amplitude traveling
waves exert a negligible Bjerknes force on bubbles, this is no longer
true for large traveling waves, and theory predicts a large Bjerknes
force, oriented in the direction of the wave propagation
\cite{kochkrefting2004,mettin2005,mettin2007}.  This means that an
ultrasonic source emitting a traveling wave would strongly repel the
bubbles nucleating on its surface.  This issue has been investigated
theoretically by Koch and co-workers \cite{kochmettin2004}, who,
assuming an arbitrary wave, traveling in the sonotrode direction and
standing in the perpendicular plane, showed that the conical bubble
structure observed under large area transducers
\cite{moussatovgrangerdubus2003,camposdubus2005} could be partially
reproduced by particle models. 

Other complex bubble structures can be observed in other
configurations, such as ultrasonic baths, and were conjectured to
result from a combination of both traveling and standing waves
\cite{mettin2005}. This raises the issue of the origin of such
traveling waves, which was one of the motivation of the present paper
and the companion one (which we will denote hereinafter by
{\firstpaper}). In the latter, we showed that traveling waves appear
as a simple consequence of the attenuation by inertial bubbles. The
model presented in {\firstpaper} constitutes therefore the missing
link in the theory, and allows to calculate the acoustic field without
any a priori on its structure, just from the knowledge of the
vibrations of the ultrasonic emitter, whatever its complexity. From
there, the Bjerknes force field can be calculated, and the shape of
the structures formed by the bubble paths in the liquid can be
examined.

\olchange{Before going further, one should remind that
  Eq.~(\ref{bjerknes}) is an over-simplification of the complex
  problem of bubbles translational motion. A correct representation of
  the latter requires to write Newton's second law for the bubble,
  accounting not only for the instantaneous driving force $-V\nabla p$
  [of which (\ref{bjerknes}) is the time-average], but also for
  viscous drag and added-mass forces \cite{magnaudeteames2000}. All
  members of such an equation are dependent of the bubble radial
  dynamics, so that considering Eq.~(\ref{bjerknes}) as a mean force
  pushing the bubbles is a reduced view of the reality, masking the
  periodic translational motion superimposed to the (macroscopically
  visible) average translational motion. This is historically
  justified, since the first studies on Bjerknes forces aimed at
  localizing the stagnation points of the bubbles where $\fbun$
  vanishes \cite{ellercrum70}, as is the case in the center of
  single-bubble levitation experiments \cite{gaitancrum92,matula99}.
  Slightly extrapolating this point of view, if one accepts that the
  average bubble velocity can be obtained approximately by a balance
  between Eq.~(\ref{bjerknes}) and an average viscous drag force, a
  terminal mean velocity of the bubble can be calculated, which
  allowed for example successful particle simulations of bubble
  structures \cite{parlitz99}. This raises the issue of nontrivial
  averaging procedure for moderate or large drivings
  \cite{reddyszeri2002bjerknes,kreftingtoilliez2006}, which may be
  performed by elaborate multiple scales procedures
  \cite{toilliezszeri2008}. However, some experimental situations
  exist where such a terminal velocity cannot be defined, and a bubble
  may wander between the nodes and antinodes of a standing wave
  \cite{khanna2003}. The description of such a phenomenon requires the
  simultaneous resolution of the instantaneous radial and
  translational equations of the bubble, initially proposed in
  Ref.~\cite{watanabekukita93}, and improved recently by a Lagrangian
  formulation \cite{doinikov2002,doinikov2005,doinikov2005review}. The
  main result of the latter studies is that the radial and
  translational motions are strongly coupled, so that the bubble
  dynamics equation is also affected by the translational motion.
  Direct simulation of the two coupled equations in standing waves
  fields reveal that, apart from the classical scheme of bubble
  migration toward stagnation points, some bubbles may have no spatial
  attractors and can wander indefinitely between a node and an
  antinode, as observed in \cite{khanna2003}. Such a behavior, termed
  as ``translationally unstable'', has been found to result from an
  hysteretic response of the radial bubble dynamics below the main
  resonance \cite{mettindoinikov2009}.  }


\olchange{In spite of the latter remarks, we will keep in this paper
  the classical picture of the mean primary Bjerknes force defined by
  Eq.~(\ref{bjerknes}) acting on the bubbles, and focus on the effects
  of traveling waves. The paper is organized as follows: in} section
2, we will first briefly recall the main results on the primary
Bjerknes force, and indicate how it can be calculated for an arbitrary
bubble dynamics in a given acoustic field.  In section 3, we will
calculate the Bjerknes force field in acoustic fields calculated with
the model proposed in {\firstpaper}, which we will briefly recall in
section~3.1. First, in section~3.2, the 1D configuration examined in
{\firstpaper} will be considered. Then, in section~3.3, we will
examine a 2D axi-symmetrical configuration, constituted by a large
area sonotrode emitting in a large bath, similar to the experiments
reported in
\olchange{Refs.~\cite{moussatovgrangerdubus2003,camposdubus2005,dubusvanhille2010}}.
Finally, section~3.4 will address another 2D configuration, mimicking
an ultrasonic bath in which the acoustic field is produced by a plate
undergoing flexural vibrations. For both 2D configurations, the bubble
paths will be drawn from the knowledge of the acoustic and Bjerknes
force fields at every point in the liquid. The structures obtained
will be compared to experimental results of the literature and
discussed.

\section{Primary Bjerknes forces}
\subsection{Intuitive analysis and linear case}
The physical origin of the primary Bjerknes force can be recalled
simply by considering a mono-dimensional wave. The instantaneous
pressure force exerted by the external liquid on a liquid sphere that
would replace the bubble is approximately the difference $\Delta p$
between the instantaneous acoustic pressures on two opposite sizes of
the sphere, multiplied by the bubble area $S$. Besides, the
pressure difference $\Delta p$ is roughly $\sdsurd{p}{x} \times 2R$,
so that the instantaneous force is roughly $ 2R\times S \times
\sdsurd{p}{x} \simeq V \sdsurd{p}{x} $. Generalizing this result in 3D
Eq.~(\ref{bjerknes}) is recovered.

\begin{figure}[ht]
  \centering
  \includegraphics[width=\linewidth]{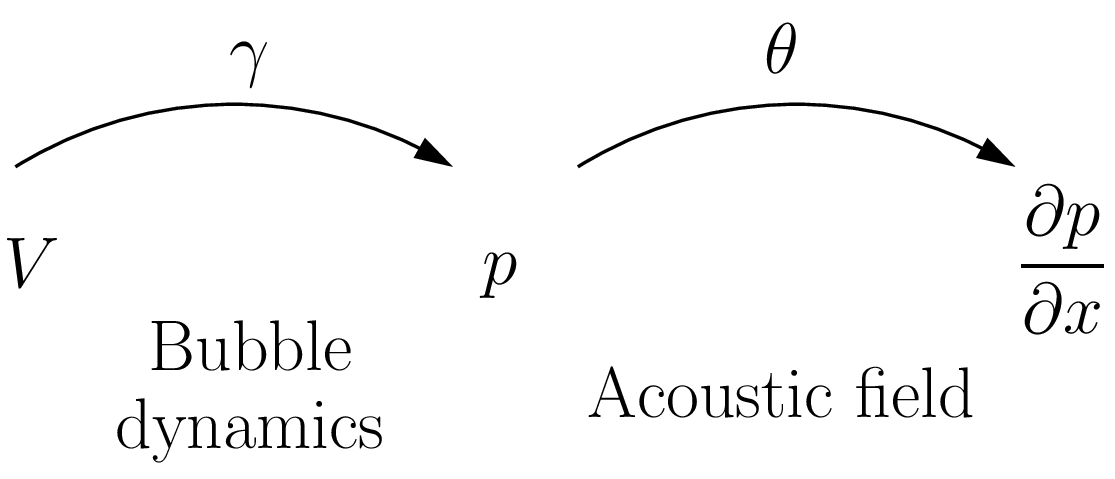}
  \caption{Schematic interpretation of Eq.~(\ref{bjerknes}). The phase
    shift between bubble volume $V$ and pressure gradient $\sdsurd{p}{x}$
    can be decomposed into two parts: $\phivp$ depending on the bubble
    dynamics, and $\phipgradp$ depending on the acoustic field.}
 \label{figpgradp}  
\end{figure}

Along an acoustic cycle, the bubble therefore wanders forward and
backward along the direction of the pressure gradient, under the
influence of this instantaneous force, but the two motions may not
exactly compensate, because the bubble may be for example larger when
the pressure gradient is directed forward than when it is directed
backward. The average force is therefore a matter of phase between the
volume $V$ and the pressure gradient $\sdsurd{p}{x}$, which can be
better understood with the schematic representation of
Fig.~\ref{figpgradp}: the phase shift between $V$ and $\sdsurd{p}{x}$
can be decomposed into two part: the phase shift $\phivp$ between
volume $V$ and pressure $p$, and the phase shift $\phipgradp$ between
$p$ and its gradient $\sdsurd{p}{x}$. The former depends on the way the bubble
responds to the local acoustic field, that is on the bubble dynamics,
whereas the latter depends on the structure of the acoustic field. All
the results mentioned in the introduction can be interpreted from this
picture.

Let's take for example the case of sub-resonant bubble oscillating
linearly. In this case, pressure and volume are in opposition
($\phivp=\pi$).  In a pure 1D linear standing wave, away from pressure
nodes or antinodes, $p$ and $\sdsurd{p}{x}$ are either in phase
($\phipgradp=0$), or in phase opposition ($\phipgradp=\pi$), depending
on the location relative to the pressure antinode.
Thus, the phase shift between $V$ and $\sdsurd{p}{x}$ is either $0$ or
$\pi$. This yields therefore a large average value for the product
$\left<V\sdsurd{p}{x} \right>$, except in the pressure antinodes and
nodes where it is zero.  Conversely, for a traveling wave, pressure
and pressure gradient are in quadrature ($\phipgradp = \pi/2$), so that
$\left<V\sdsurd{p}{x} \right>$ is clearly zero in this case.

If the analysis is rather simple for linear oscillations, this is
no longer the case for strongly nonlinear inertial oscillations. In
that case, the bubble radial motion is mainly driven by the inertia of
the liquid, and the bubble radius contains a large out-of-phase
component with respect to the driving pressure $p$.  This is the
reason why pressure antinodes may become repulsive even for a
sub-resonant bubble in a large amplitude standing wave
\cite{akhatov97une}, and why the Bjerknes force may become very large
in traveling waves \cite{kochkrefting2004,kochmettin2004}. The next
section quantifies this qualitative analysis

\subsection{General calculation of the Bjerknes force}
The model described in {\firstpaper} was shown to result from the
assumption that the bubbles mainly respond to the first harmonic of
the field, which we termed as ``first harmonic approximation'' (FHA).
We therefore assume that the pressure field in the liquid is
mono-harmonic at angular frequency $\omega$, and defined in any point
$\vvr$ by
\begin{equation}
  \label{defphar}
  \pact(\vr,t) = p_0 + \undemi \biggl( 
  P(\vr)e^{i\omega t} + \conj{P}(\vr)e^{-i\omega t} 
  \biggr),
\end{equation}
which, writing $P =|P(\vr)|\exp\bigl(i\phi(\vr)\bigr)$, can be recast
as
\begin{equation}
  \label{defpaco}
  \pact (\vvr,t) = p_0 + |\pac (\vvr)| \cos
  \left[\omega t+\phi(\vvr) \right]
\end{equation}
This expression may represent a traveling wave, a standing wave, or
any combination of both. We also define the pressure gradient in
general form as
\begin{equation}
  \label{defpgrad}
  \pgradt  (\vvr,t) = \pgrad (\vvr) \cos
  \left[\omega t + \psi_i(\vvr) \right],
\end{equation}
where the fields  $\pgrad$ and $\psi_i$ can be expressed as functions
of $\pac$ and $\phi$ once the acoustic field is known. 

The following two extreme cases deserve special consideration: for a
standing wave, $\phi(\vvr)=\phi_0$, so that
$\pgrad(\vvr)=\sdsurd{\pac}{\ri}$ and $\psi_i(\vvr)=\phi_0$; for a
traveling wave, $\pac(\vvr)=P_0$ and $\phi(\vvr)=-\vv{k}.\vvr$ so that
$\pgrad(\vvr) = k_i P_0$ and $\psi_i(\vvr)=\phi(\vvr)-\pi/2$.

The expression of the Bjerknes force on the bubble located at $\vr$
reads, from (\ref{bjerknes}): 
\begin{equation}
  \label{defbjerknesint}
  \fbun_i = -\pgrad (\vvr) \frac{1}{T}\int_0^{T} V(\vr, t)  \cos
  \left[\omega t + \psi_i(\vvr) \right]  \; d t,
\end{equation}
where $T$ is the acoustic period and $V(\vr,t)$ is the instantaneous
volume of a bubble located at $\vr$ and can be calculated by solving a
radial dynamics equation, for example:
\begin{equation}
  \label{rayleigh}
  \rho_l\left(R\ddot{R} + \frac{3}{2}\dot{R}^2 \right) = 
  p_g-\frac{2\sigma}{R}-4\mu_l\frac{\dot{R}}{R} - p(\vr,t).
\end{equation}
The bubble volume depends on $\vr$ because two bubbles located at
different points may be excited by fields of different amplitudes
$|P|$ but also different phases $\phi$.  However, in order to be able
to calculate the volume $V$ of any bubble over one acoustic period
independently of its spatial location, we must fix the phase of the
driving field in Eq.~(\ref{rayleigh}) by a convenient change of
variables. We therefore set, taking this opportunity to
non-dimensionalize the variables:
\begin{equation}
  \label{padimrt}
  \pact(\vvr,t) = p_0\left( 1 - |P^*| \cos \tshift \right),
\end{equation}
where the minus sign has been chosen to be consistent with earlier
studies \cite{hilgenbrennergrosslohse98,louisnard2008}, and, comparing
this expression with Eq.~(\ref{defpaco}), we get:
\begin{gather}
  \label{defp}
  P^* = \pac(\vvr)/p_0, \\
  \label{defx}
  \tshift(\vr,t) = \omega t + \phi(\vvr) - \pi.
\end{gather}
We now note $V(\tau^*)$ the volume of the bubble when it is driven by
the pressure field (\ref{padimrt}), and making the change of variables
in (\ref{defbjerknesint}), we get:
\begin{equation}
  \label{defbjerknesintbis}
  \fbun_i = \pgrad (\vvr) \frac{1}{2\pi}\int_0^{2\pi} V(\tshift)  \cos
  \left[\tshift -\phi(\vr)+ \psi_i(\vvr) \right]  \; d\tshift,  
\end{equation}
which we recast as:
\begin{eqnarray}
  \nonumber
  \fbun_i &=& \pgrad (\vvr) 
  \biggl(
  I_C \cos \left[ \phi(\vr) - \psi_i(\vvr) \right] \\
  \label{defbjerknesintter}
  \quad & &+ 
  I_S \sin \left[ \phi(\vr) - \psi_i(\vvr) \right]
  \biggr ).,
\end{eqnarray}
with
\begin{gather}
  \label{defIc}
  I_C = \frac{1}{2\pi}\int_0^{2\pi} V(\tshift)  \cos \tshift \; d\tshift, \\
  I_S = \frac{1}{2\pi}\int_0^{2\pi} V(\tshift)  \sin \tshift \; d\tshift.
\end{gather}
This decomposition (which was suggested in a slightly different form
by Mettin \cite{mettin2005}), has the advantage to clearly decompose
the respective influences of the bubble dynamics through integrals
$I_C$ and $I_S$, and the one of the acoustic field, through the phase
shift $\phi(\vr) - \psi_i(\vvr)$. In the case of a 1D wave, the latter
corresponds to the angle $\theta$ that we defined in figure
\ref{figpgradp}. Integral $I_C$ measures the standing wave
contribution to the Bjerknes force ($\phi-\psi = 0$ or $\pi$), while
integral $I_S$ measures the traveling wave contribution ($\phi-\psi =
\pm\pi/2$).

The two integrals $I_C$ and $I_S$ can be easily calculated numerically
by solving a radial dynamics equation to obtain $V$ and averaging over
one period.  They can also be calculated analytically, trivially for
linear oscillations, and in a more complex manner for inertial
oscillations in a small size range above the Blake threshold
\cite{louisnard2008}.  The correct matching between the two extreme
cases is however difficult, so that we will use the numerical values
hereinafter. 

\begin{figure}[ht]
  \centering
  \includegraphics[width=\linewidth]{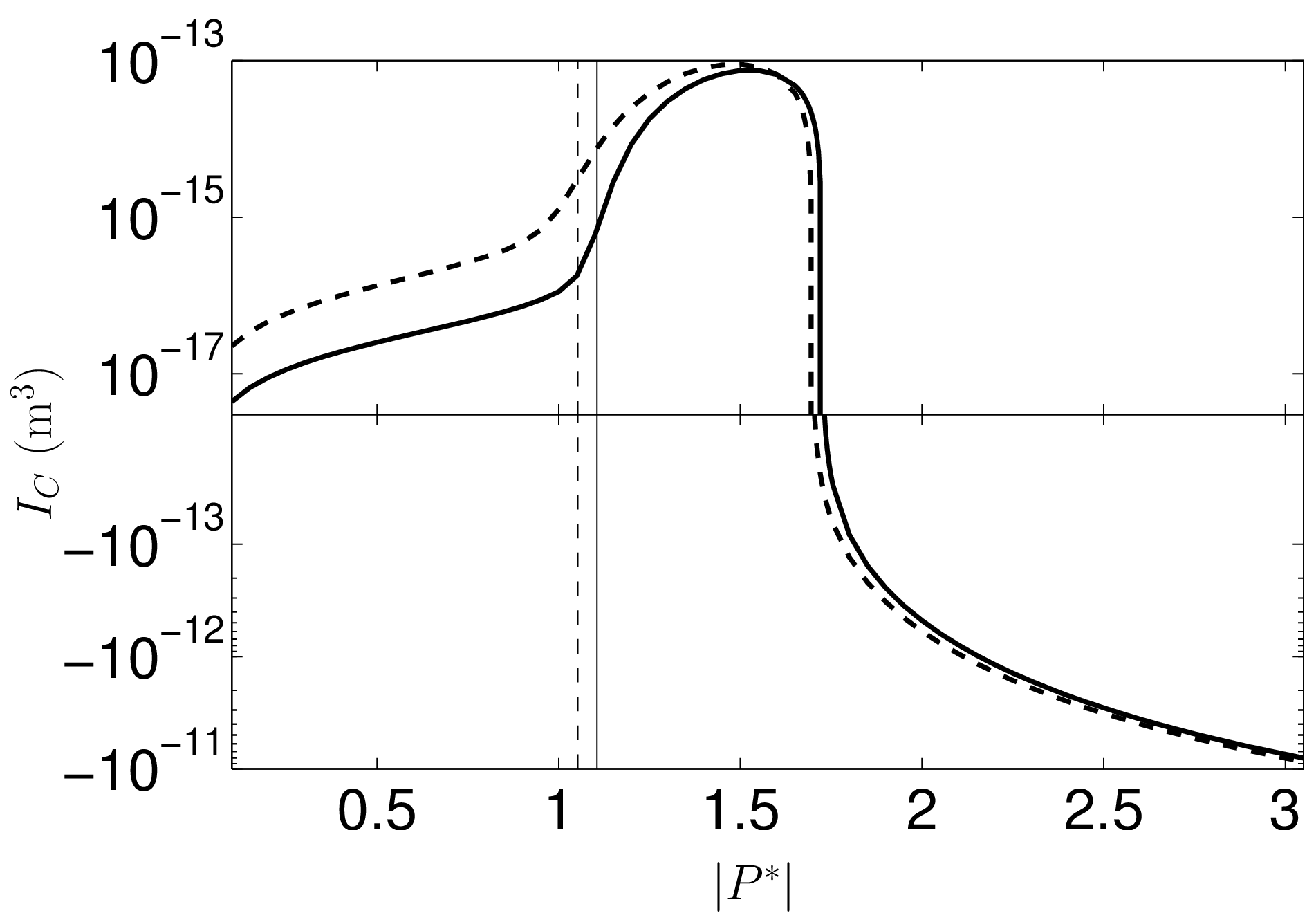}
  \includegraphics[width=\linewidth]{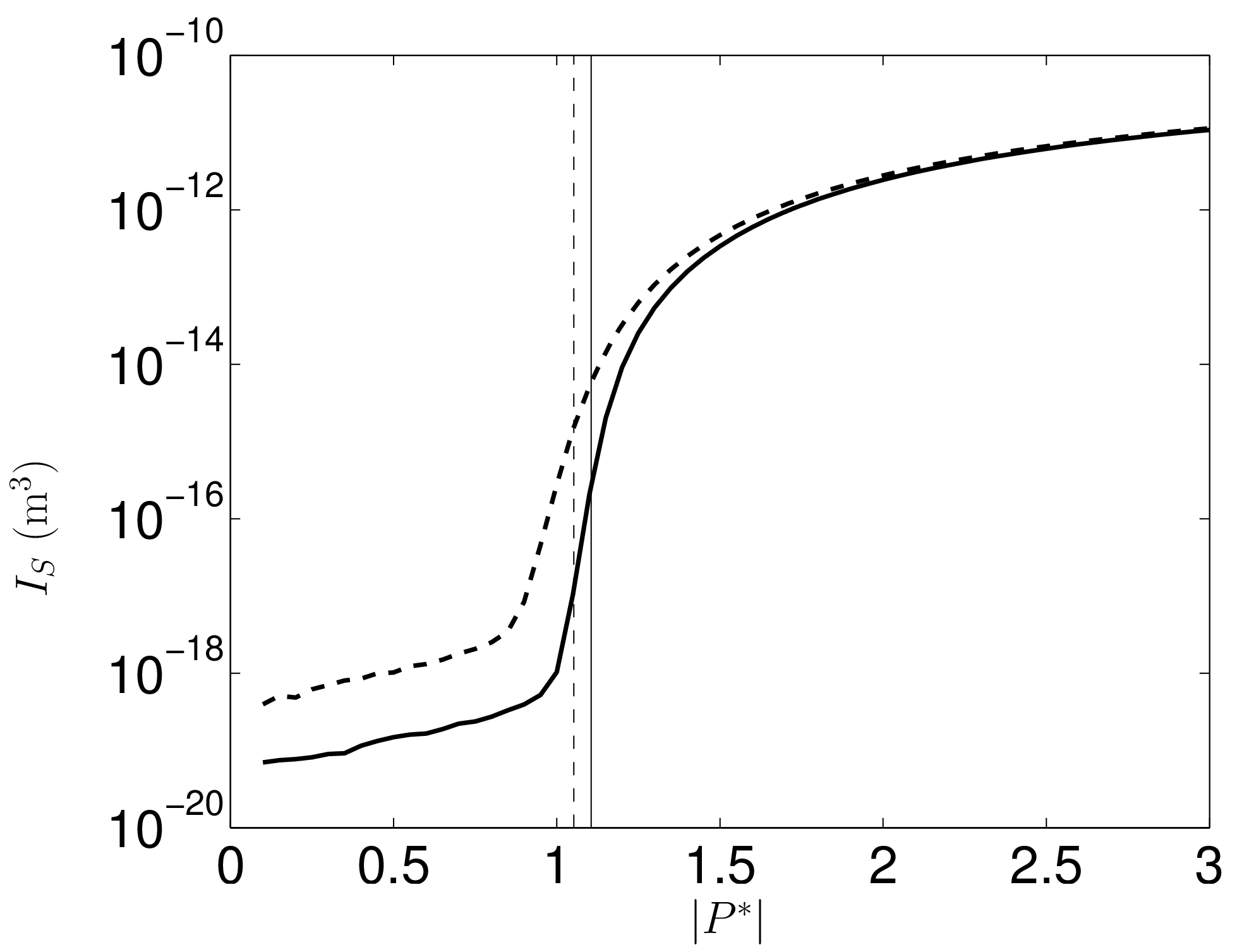}
  \caption{Average quantities $I_C$ and $I_S$ for an air bubble of
    ambient radius 3~$\mu$m (solid line) and 5~$\mu$m (dashed line) in
    water, as a function of the dimensionless acoustic pressure
    $|P^*|$ for frequency of 20 kHz. \olchange{The vertical lines
      represent the Blake thresholds for 3~$\mu$m (solid) and 5~$\mu$m
      bubbles (dashed).}}
 \label{figIcIs}  
\end{figure}

The results are displayed in Fig.~\ref{figIcIs} in the case of air
bubbles in water in ambient conditions for two ambient radii $R_0=3$
$\mu$m (solid lines) and $R_0=5$ $\mu$m (dashed lines). The integral
$I_C$ is represented in signed logarithmic scale (note the different
scales for the positive and negative parts). It is seen that it is
positive for low drivings, and quickly increases near the Blake
threshold by about 3 orders of magnitude. In this range of acoustic
pressures, antinodes are attractive. Then, above $|P^*|=1.7$ bar, $I_c$
becomes largely negative and the antinodes become repulsive, as was
found in Ref.~\cite{akhatov97une}. 

The integral $I_S$ is always positive, is very weak below the Blake
threshold, and drastically increases above the Blake threshold (by 6 to
7 orders of magnitude). This predicts that, as reported in
Ref.~\cite{kochkrefting2004}, the Bjerknes force can become very large in
traveling waves.





\section{Results}
\subsection{Simulation method}
The complex acoustic field $P$ is obtained by solving a nonlinear
Helmholtz equation, which has been detailed in {\firstpaper} and is briefly
recalled here for completeness:
\begin{equation}
  \label{helmholtzNL}
  \nabla^2 P + k^2\left(|P| \right) P = 0.
\end{equation}
where the complex wave number is given by:
\begin{gather}
  \label{defk2r}
  \Re(k^2) =  \frac{\omega^2}{c_l^2} + \frac{4\pi R_0 \omega^2 N}
  {\omega_0^2-\omega^2},\\
  \label{defk2i}
  \Im\left(k^2 \right) = - 2\rho_l\omega N \frac{\pith+\piv}{|P|^2}.
\end{gather}
The bubble number $N$ is defined as a step function: it is assumed
zero in the zones where the acoustic pressure is less than the Blake
threshold, and is assigned to a constant value in the opposite case. 
\begin{equation}
  \label{choixN}
  N = \left\{
    \begin{array}{lll}
      N_0 & \text{if } |P| > P_B \\
      0 & \text{if } |P| < P_B 
    \end{array}
 \right.
\end{equation}
%

\subsection{1D results}
\begin{figure}[ht]
  \centering
  \includegraphics[width=\linewidth]{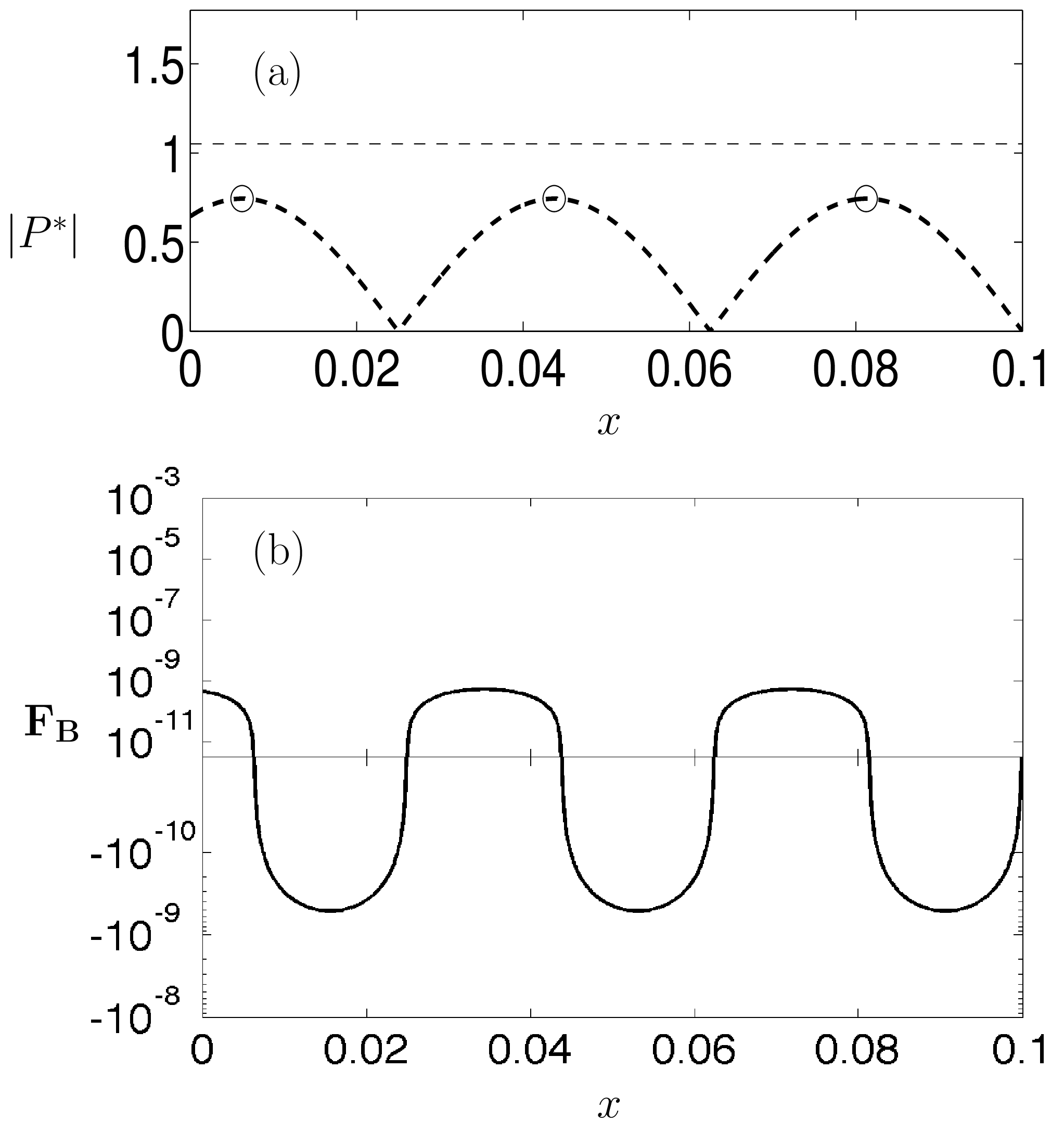}
  \caption{Simulation of the sound field and resulting primary
    Bjerknes force for 5~$\mu$m air bubbles in water, with an emitter
    displacement amplitude is $U_0$= 0.3 $\mu$m at 20~kHz.  (a):
    Pressure amplitude profile. \olchange{The horizontal dashed line
      represents the Blake threshold for 5 $\mu$m bubbles.} (b):
    Bjerknes force exerted on bubbles at each point of the domain
    (with signed logarithmic scale in ordinate). The circles in Fig.
    (a) represents the stable stagnation points for the bubble.}
 \label{figbjerkneslow}
\end{figure}

We first calculate the primary Bjerknes force for the same 1D
configuration as in {\firstpaper}: the domain length is 10~cm, the
right boundary is assumed infinitely soft, air bubbles of ambient
radius 5~$\mu$m in water. For a low amplitude of the emitter
($U_0=0.3$~$\mu$m), the pressure amplitude profile is almost a perfect
standing wave (Fig.~\ref{figbjerkneslow}a), and
Fig.~\ref{figbjerkneslow}b exhibits the classical picture of a
somewhat low primary Bjerknes force pushing the bubbles towards the
antinodes (note that the distortion of the force profile in
Fig.~\ref{figbjerkneslow}b is an artifact of the signed logarithmic
scale used in ordinate). In this case, the value of the Bjerknes force
mainly owes to the $I_C$ term in Eq.~(\ref{defbjerknesintter}).

\begin{figure}[ht]
  \centering
  \includegraphics[width=\linewidth]{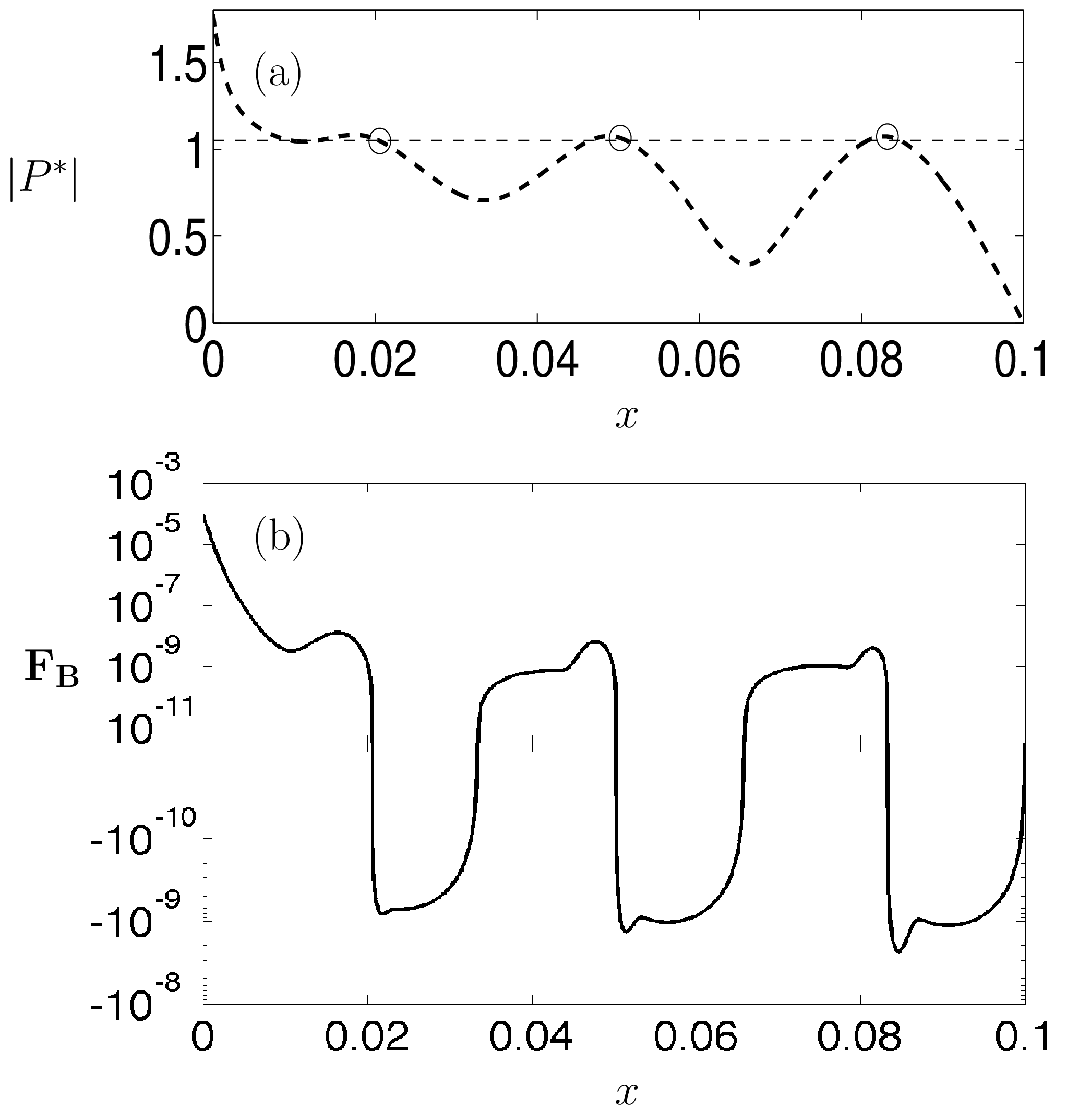}
  \caption{Simulation of the sound field and resulting primary
    Bjerknes force for 5~$\mu$m air bubbles in water, with an emitter
    displacement amplitude is \olchange{$U_0$= 5 $\mu$m} at 20~kHz.
    (a): Pressure amplitude profile. \olchange{The horizontal dashed
      line represents the Blake threshold for 5 $\mu$m bubbles.} (b):
    Bjerknes force exerted on bubbles at each point of the domain
    (with signed logarithmic scale in ordinate). The circles in Fig.
    (a) represents the stable stagnation points for the bubble.}
 \label{figbjerkneshigh}
\end{figure}

For larger emitter amplitude (\olchange{$U_0=5$~$\mu$m}), the pressure amplitude
profile is strongly damped near the emitter, as already commented in
{\firstpaper} (Fig.~\ref{figbjerkneshigh}a).  This strong attenuation
produces a noticeable traveling part in the wave and thus a large
$I_S$ term in Eq.~(\ref{defbjerknesintter}), which, as shown in
Fig.~\ref{figbjerkneshigh}b, results in a huge positive force near the
emitter almost 6 orders of magnitude higher than the maximal force
visible in Fig~\ref{figbjerkneslow}b (note that the scales used in
both figures are identical).  The bubbles would consequently be
strongly expelled from the emitter, and travel right to the first
stagnation point which is located somewhat far from the emitter (see
the leftmost circle marker on Fig.~\ref{figbjerkneshigh}a). The
occurrence of such stagnation points has been proposed to be the key
mechanism for bubble conical structure formation
\cite{kochmettin2004,mettin2005}. We will show in the next section
that this is partially true, but involves some additional subtleties.

\subsection{Sonotrode}
\begin{figure}[ht]
  \centering
  \includegraphics[width=\linewidth]{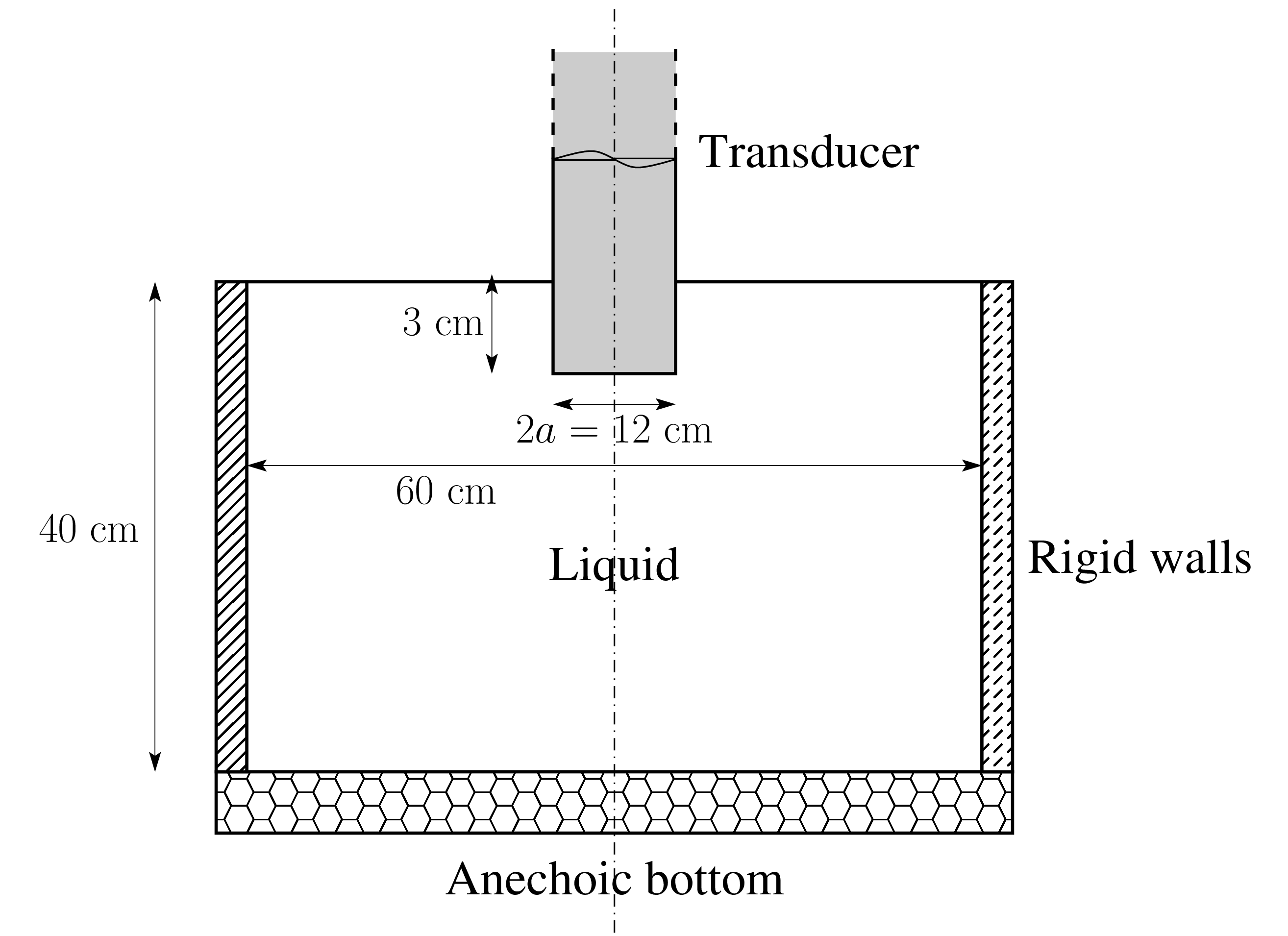}
  \caption{Axi-symmetrical geometry for large area sonotrode.}
 \label{figgeomcone}  
\end{figure}

\olchange{\subsubsection{Experiments with a 12~cm diameter sonotrode}}
In this section we compare the results of our model to cone bubble
structures images presented in Ref.~\cite{moussatovgrangerdubus2003}.
The latter work used a 100~cm $\times$ 60~cm rectangular water tank of
40~cm depth, where a sonotrode of diameter $2a=$~12~cm driven at
20.7~kHz is immersed at 3~cm below the free liquid level
(Fig.~\ref{figgeomcone}).  

The geometry simulated follows the experimental configuration used in
Ref.~\cite{moussatovgrangerdubus2003} as closely as possible. However,
in order to avoid time-consuming 3D simulations, we replaced the
rectangular tank by a cylindrical one with the same depth and a
diameter of 60~cm, and simulate only a half-plane cut in
axi-symmetrical mode.  The characteristic of the bottom of the tank is
not specified in Ref.~\cite{moussatovgrangerdubus2003} but following a
similar work by the same authors in which the same bubble structures
were observed \cite{camposdubus2005}, we considered an anechoic tank
bottom. The lateral sides of the tank were taken as infinitely rigid
boundaries, and the free liquid surface as infinitely soft. The
nonlinear Helmholtz equation with the latter boundary conditions was
solved in axi-symmetrical geometry, using the commercial COMSOL
software.

The transducer is also simulated in order to account for its lateral
deformation in the liquid, which, as will be seen below, is necessary
to catch some experimental features. However, since details on the
internal structure of the transducer are not given in
Refs.~\cite{moussatovgrangerdubus2003,camposdubus2005}, we assumed the
latter made of steel and following a non-dissipative elastic behavior
represented by Hooke's law (with Young modulus $E=2\times 10^{11}$ Pa,
Poisson ratio $\nu=0.3$, density $\rho_S = 7900$~kg/m$^3$).  The
vibration of the transducer is coupled to the acoustic field in the
liquid by using the convenient cinematic and dynamic interface
conditions, as detailed in Ref.~\cite{louisnardgonzalez2009}.  We
simulate only the bottom part of the transducer, containing the whole
part immersed in the liquid and an arbitrary small length (3~cm) of
the emerged part.  A uniform sinusoidal displacement of amplitude
$U_0$ is imposed on the upper boundary of the simulated transducer
rod. In order to match the conditions of the experiments,  the
acoustic intensity $I$ entering the medium through the lower boundary of
the sonotrode is calculated by: \cite{louisnardgonzalez2009}
\begin{equation}
  \label{intensity}
  I = \frac{1}{\pi a^2}\iint\limits_S \frac{1}{2}\Re\left(P V^* \right)\;d S
\end{equation}
Both parameters $U_0$ and $N_0$ are varied in order to obtain the
required value of $I$. 

In order to tentatively exhibit the bubble structures formed in a
given configuration, the bubble paths, generally termed as
``streamers'', will be materialized by drawing the "streamlines" of
the Bjerknes force field in some parts of the liquid.  The adequate
choice of the starting points of these streamlines is difficult,
because it would require a clear knowledge of the bubble nucleation
process. Solid boundaries are known to act as sources of bubble
nuclei, where the latter may be trapped by microscopic crevices
\cite{crum82}. Common observation of cavitation experiments indeed
show that bubbles often originate from the transducer area, which
might suggest that the release of crevice-trapped bubbles is more
efficient on vibrating surfaces. We will therefore launch
systematically streamlines of the Bjerknes force field from
equidistant points of vibrating boundaries, and we will term them as
"\sstreamers{}".

Besides, many bubble structures appear far from solid boundaries as a
more or less complex set of bubble filaments \cite{mettin2005}. In
that case, bubble seem to originate from given points of the bulk
liquid, but the precise mechanism of nucleation of such bubbles is not
clear. Although it has long been thought that sub-micronic nuclei could
grow up to the Blake threshold by rectified diffusion
\cite{neppirasphysrep,leightonacousticbubble}, this is ruled out by
nonlinear theory, since a sub-Blake bubble cannot grow by rectified
diffusion~\cite{louisnardgomez2003}. Coalescence is an alternative
growth process, but this issue is yet unresolved. We will therefore
assume that a bubble is visible and contributes to structures only if
it is inertially oscillating, that is in zones above the Blake
threshold.  This is anyway consistent with our assumption on the
bubble density (\ref{choixN}) used to calculate the acoustic field. We
will therefore launch streamlines from arbitrary points located on the
calculated curves $|P|=P_B$, where $P_B$ is the Blake threshold
($P_B=1.178$~bar for 2~$\mu$m bubbles in ambient conditions, see
{\firstpaper}). We will refer to such streamlines as ``\lstreamers{}''
and we will represent then with a color different from surface
streamlines in order to distinguish them.

\begin{figure}[ht]
  \centering
  \includegraphics[width=0.7\linewidth]{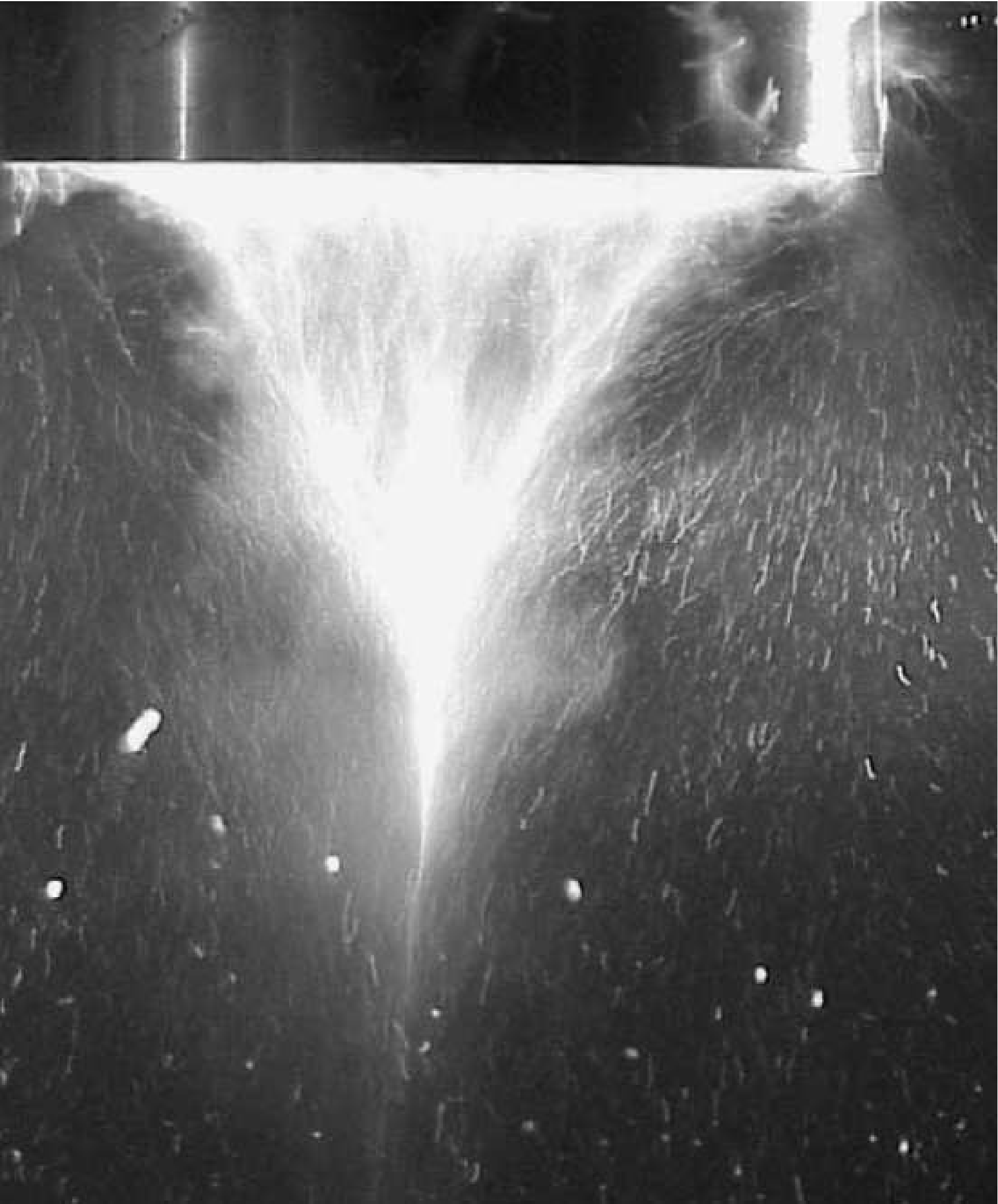}
  \caption{Original image of cone bubble structure (reprinted from
    Ref.~\cite{moussatovgrangerdubus2003}, with permission from
    Elsevier).  The transducer diameter is 12~cm.}
  \label{figConeComplet} \end{figure}

Figure~\ref{figConeComplet} displays one of the original images
obtained in Ref.~\cite{moussatovgrangerdubus2003} for an acoustic
intensity $I=8.2$~W. The cone is completely formed and ends in a long
tail undergoing lateral fluctuations, which explains the slightly
non-symmetric shape of the structure. Besides, it can be easily seen
that a large region inside the cone is poorly populated in bubbles,
compared to the the immediate vicinity of the transducer and the
lateral boundaries of the cone. 


\begin{figure}[h!t]
  \centering
  \includegraphics[width=0.9\linewidth]{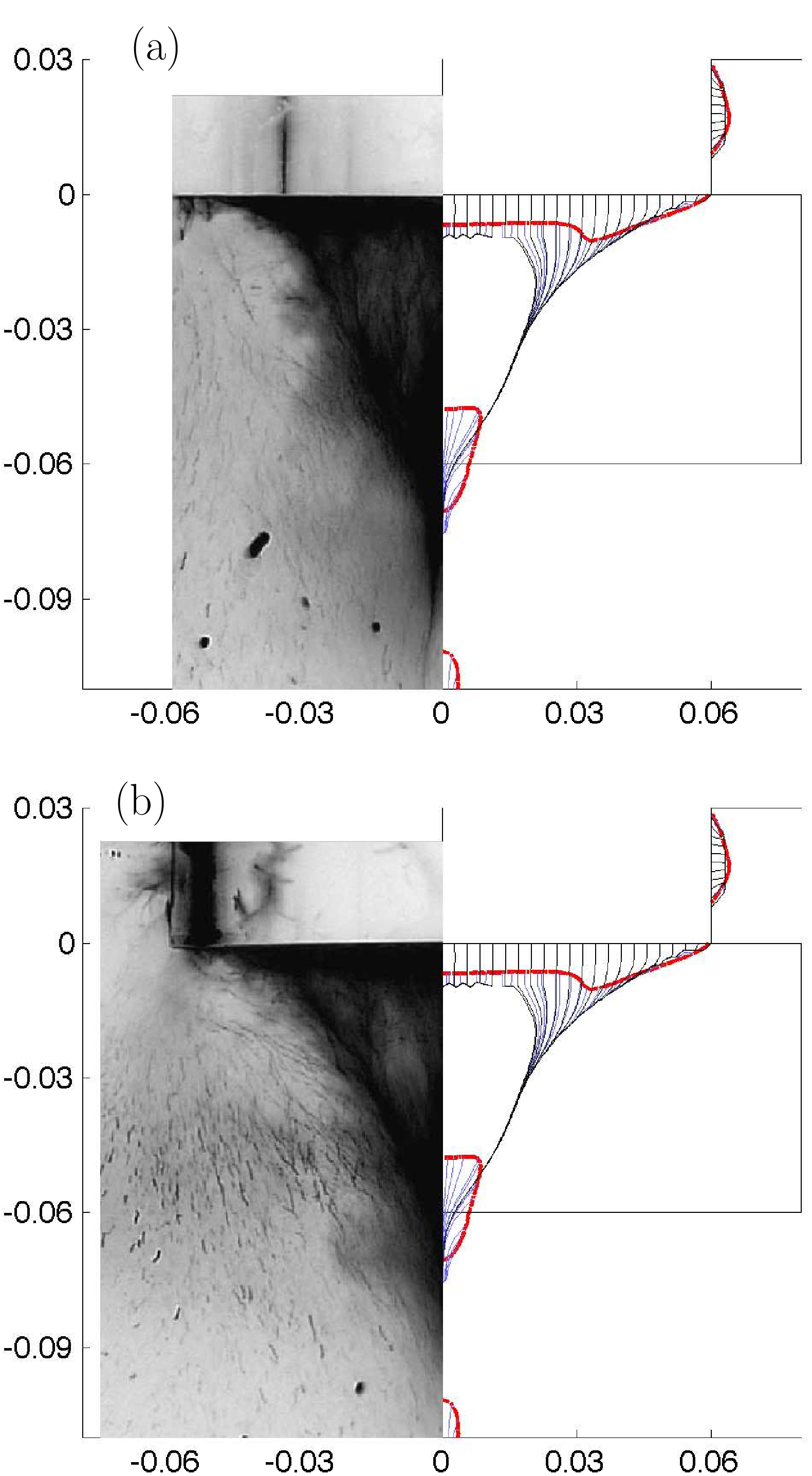}
  \caption{Comparison between the experimental picture of
    Fig.~\ref{figConeComplet} (the image of
    Ref~\cite{moussatovgrangerdubus2003} has been video reversed in
    order to make the comparison easier) and the numerical results
    obtained for 2~$\mu$m air bubbles, with $U_0=1.4$~$\mu$m,
    \olchange{$N_0=360$~bubbles.mm$^{-3}$}. The input intensity is
    8.2~W.cm$^{-2}$. The black lines are the \sstreamers{}, the blue
    lines are the \lstreamers{}, and the thick solid red line is the
    Blake threshold contour curve ($|P^*| = 1.178$ for 2~$\mu$m
    bubbles). (a) comparison with the left part of
    Fig.~\ref{figConeComplet} ; (b) comparison with the mirrored
    right part of Fig.~\ref{figConeComplet}. The horizontal and
    vertical lines in the liquid just mark the separation between
    various subdomains and do not have any physical meaning.}
  \label{figcone}
\end{figure}

We present in Fig.~\ref{figcone} a comparison between this picture and
the result of our model for $U_0=1.4$~$\mu$m and
\olchange{$N_0=360$~bubbles.mm$^{-3}$}. Because the original picture is
non-symmetric, and to ensure a maximal objectivity in the comparison,
we present our result compared both to the left part of the cone
(Fig.~\ref{figcone}a) and to its mirrored right part
(Fig.~\ref{figcone}b). Besides, the original picture has been
video-reversed in order to obtain black bubble paths on a white
background. We emphasize that this was the only image treatment
performed. The Blake threshold contour curve is displayed in thick
solid red line. The \sstreamers{}, originating from the transducer,
are displayed in black, while the \lstreamers{}, originating from
arbitrary points on the Blake threshold contour line are displayed in
blue.

First, it is seen that the global shape of the cone is correctly
reproduced. The acoustic pressure on the axial point of the emitter is
2.16~bar. Cavitation occurring near the sonotrode dissipates a lot of
energy, which produces a strong attenuation and therefore a large
traveling wave contribution in the vertical direction. Bubbles
originating from the transducer (black lines in Fig.~\ref{figcone})
are therefore strongly expelled from the sonotrode surface, the
mechanism being the same as the one explained above for 1D waves (see
Fig.~\ref{figbjerkneshigh}).

\begin{figure}[h!t]
  \centering
  \includegraphics[width=\linewidth]{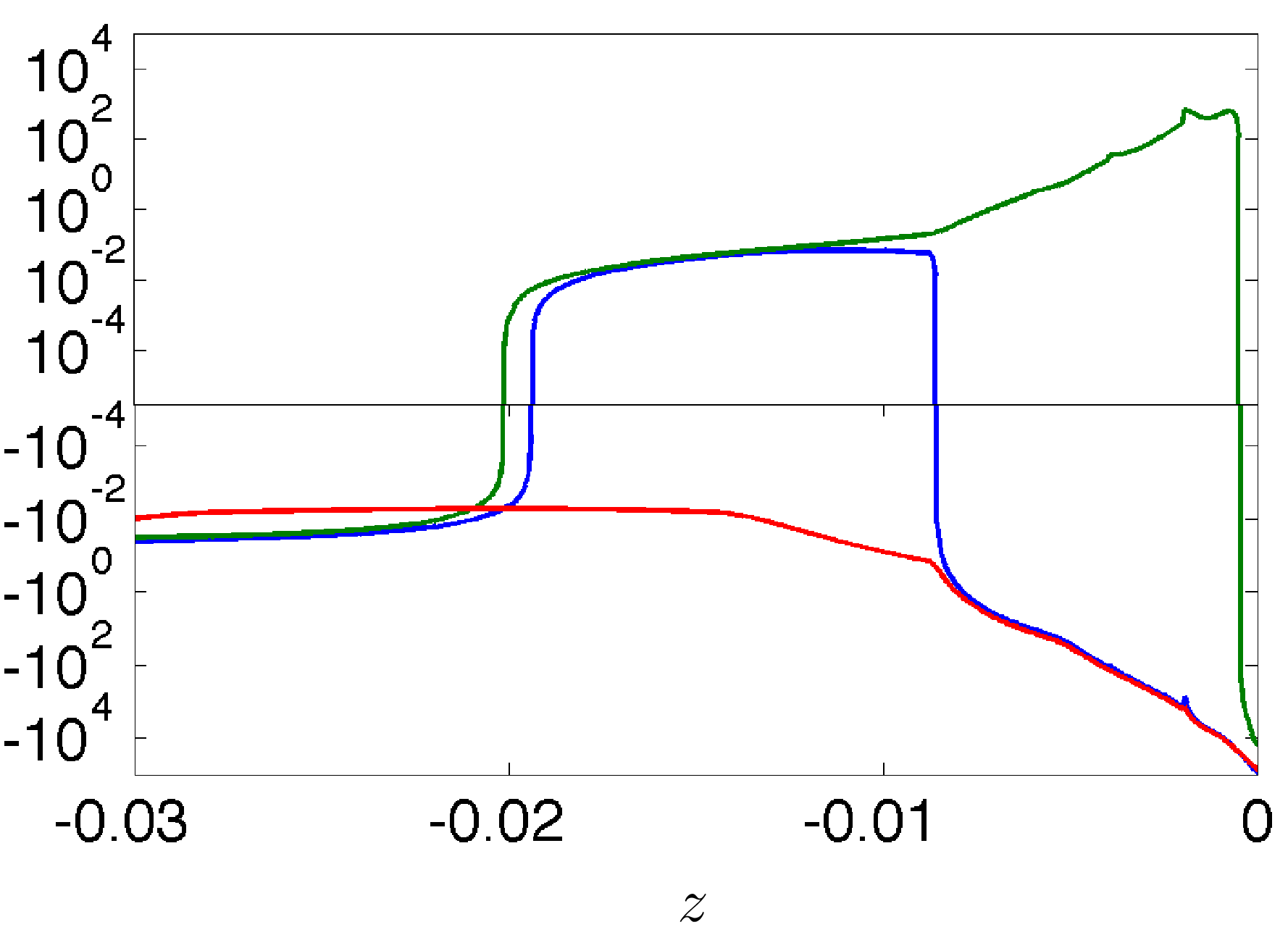}
  \caption{Blue line: magnitude of the parenthesis of $\fbun_z$ in
    Eq.~(\ref{defbjerknesintter}) along the symmetry axis.  The
    emitter is on the right of the graph. Green line: $\cos$ term in
    the parenthesis of $\fbun_z$ in Eq.~(\ref{defbjerknesintter})
    (standing wave contribution). Red line: $\sin$ term in
    the parenthesis of $\fbun_z$ in Eq.~(\ref{defbjerknesintter})
    (traveling wave contribution).}
  \label{figbjerknescomp}
\end{figure}

A more detailed analysis of the vertical component of the Bjerknes
force can be made by referring to the $z$-projection of
Eq.~(\ref{defbjerknesintter}). The green line in
Fig.~\ref{figbjerknescomp} represents the standing wave contribution
$I_C\cos\left[ \phi(\vr) - \psi_z(\vvr) \right]$ as a function of the
distance to the sonotrode (the latter being located on the right of
the graph), the red line is the traveling wave contribution
$I_S\sin\left[ \phi(\vr) - \psi_z(\vvr) \right]$, and the blue line is
the sum of the two latter. The sign of the blue line represents
therefore the sign of the $z$-component of $\fbun$, which, if
negative, corresponds to a downward oriented force. It can be seen that
near the sonotrode, the vertical Bjerknes force is dominated by the
strong repulsive traveling wave contribution, as was the case for the
1D simulation, which is clearly due to the strong attenuation of the
wave near the sonotrode. It can be noticed by the way that the
standing wave contribution is repulsive only in a small layer near the
sonotrode because the acoustic pressure is larger than the threshold
1.7~bar in this zone. Then, slightly before $z=-0.01$~m, the positive
standing wave contribution cancels exactly the traveling wave one, and
becomes dominant, so that the Bjerknes force becomes positive. This
means that, as far as only the $z$-component of the Bjerknes force is
concerned, there is a stagnation point for bubbles near $z=-0.01$~m.
In fact as will be seen below, this point is a saddle-point since the
axis is repulsive in the radial direction at this point, so that the
bubbles expelled from the sonotrode brake here in the $z$-direction
and follow their motion radially, which explains the formation of the
void region in the core of the cone. Near $z=-0.02$~m, the standing
wave contribution changes sign again, so that the force becomes
downwards and dominated by the traveling wave again. Drawing the same
curves for larger distances from the sonotrode would show that this is
the case up to another sign-change of the standing wave contribution,
near $z = -0.155$~m, which constitutes a real stagnation point since
there, the radial force is oriented towards the axis. Contrarily to
earlier interpretations \cite{kochmettin2004,mettin2005}, the present
results suggest that the tip of the cone (near $z=-0.07$~m) is not a
stagnation point, and that the real one is located well below, so that
the bubbles follow closely the axis up to the latter on an appreciable
distance. This is consistent, at least qualitatively, with the
original picture of the cone Fig.~(\ref{figConeComplet}) which shows
that the cone ends into a long fluctuating tail.

\begin{figure}[h!t]
  \centering
  \includegraphics[width=\linewidth]{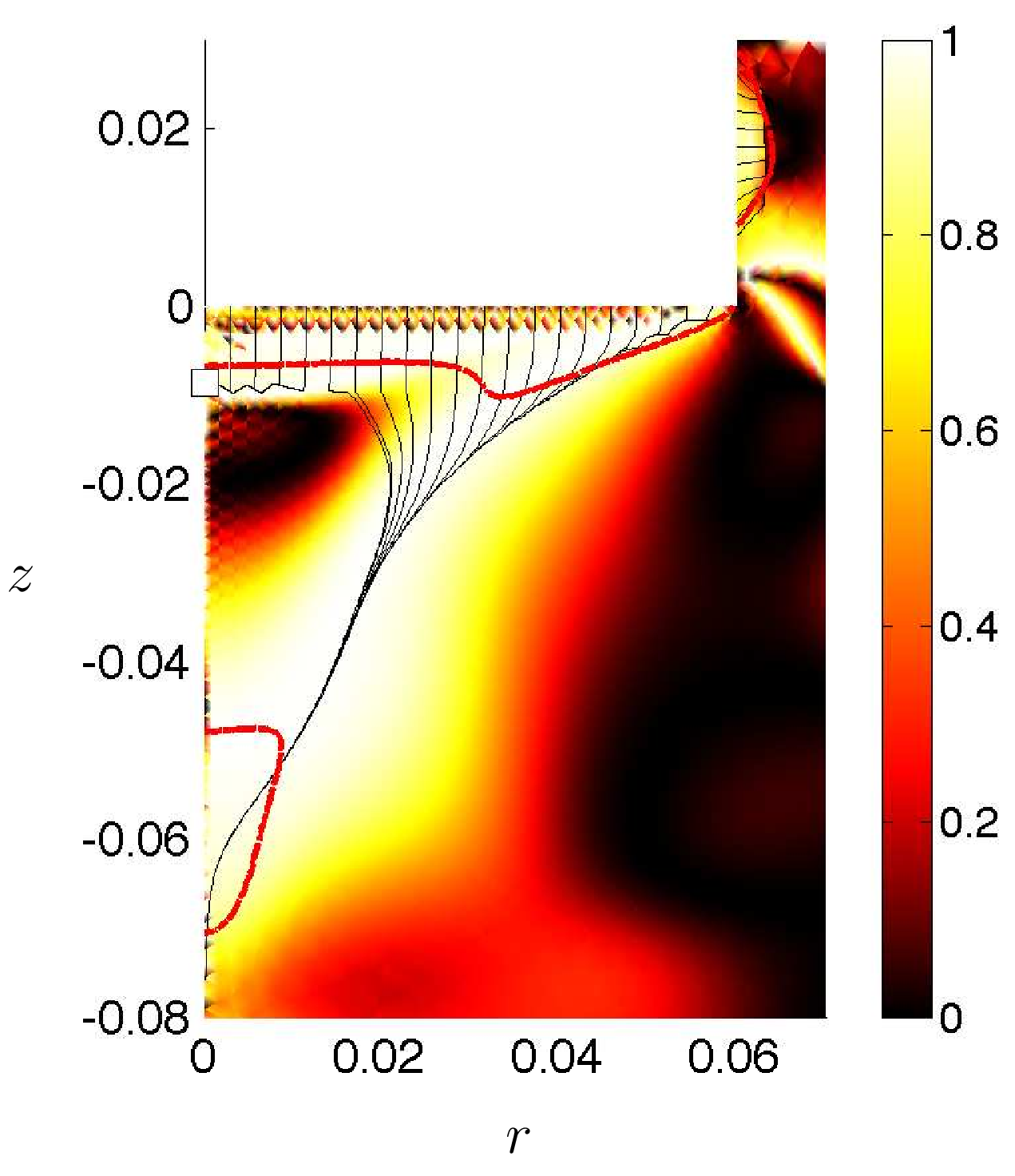}
  \caption{Color plot of $\sin^2\theta_r$, where $\theta_r$ represents
    the phase between $p$ and $\sdsurd{p}{r}$. The {\sstreamers} are
    recalled in black lines.}
  \label{figconethetar}
\end{figure}

\begin{figure}[ht]
  \centering
  \includegraphics[width=\linewidth]{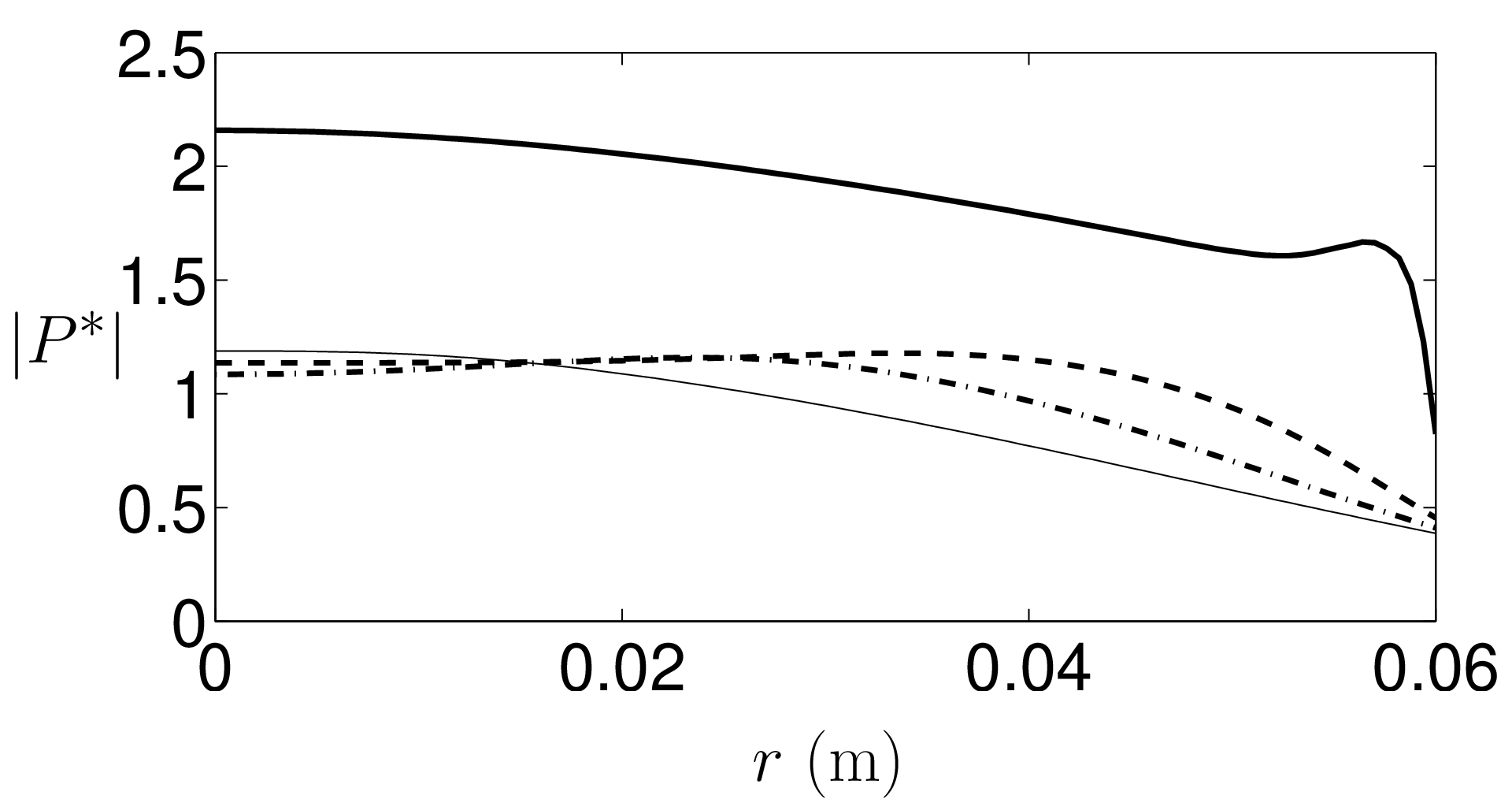}
  \caption{Radial pressure profiles at $z = 0$~m (thick solid line),
    $z=-0.01$~m(dashed line), $z=-0.02$~m (dash-dotted line),
    $z=-0.04$~m (thin solid line).}
  \label{figconeprofpa}
\end{figure}

We now look at the behavior in the radial direction.
Figure~\ref{figconethetar} displays $\sin^2\theta_r$, where $\theta_r
= \phi(\vr) - \psi_r(\vvr)$ is the phase shift between $p$ and
$\sdsurd{p}{r}$. It can be seen that a large region of traveling wave
in the r-direction ($\sin^2\theta_r \simeq 1$) surrounds approximately
the cone boundary, but that $\sin^2\theta_r$ decreases back to zero
when either entering the core of the cone, or moving outward
perpendicular to the cone boundary.  In the latter two regions, the
wave has therefore a larger standing part, so that the classical
picture of attraction by pressure antinodes and repulsion by pressure
nodes applies.  Thus, when the pressure is maximal on the axis,
bubbles converge toward the latter, and this explains the formation of
the narrow cone tip. This is the case for $z=-0.04$~m (thin solid line
in Fig.~\ref{figconeprofpa}). The opposite holds in the core of the
cone, where the variations of the acoustic pressure in the radial
direction presents a local minimum on the axis, for example at
$z=-0.01$~m (dashed line in Fig.~\ref{figconeprofpa}) or $z=-0.02$~m
(dash-dotted line).  The radial component of the Bjerknes force in
this zone is therefore oriented outwards. This is why, as mentioned
above, the point on the axis (near $z=-0.01$~m) where the
$z$-component of the Bjerknes force change sign is in fact a
saddle-point, which locally pushes the bubbles far from the axis, and
produces a void region in the heart of the cone, clearly visible on
the experimental picture.  This feature has been commented in
Ref.~\cite{moussatovgrangerdubus2003} and was attributed to the
nonlinear reversal of the Bjerknes force in standing waves near 1.7 bar.  Our
results suggest that this is not the case, and that the void region
results from a combination of a canceling $z$-component of the
Bjerknes force and a local inversion of the radial standing wave
pressure profile.

The shape of our predicted void region shows reasonable agreement with
the experiments. Furthermore, the experimental cone tip seems to be
more dense than its core. This may be due to nucleation of bubbles at
the Blake threshold in this part, as suggested by the \lstreamers{}
starting from the Blake contour loop just above the cone tip in
Fig.~\ref{figcone} (blue lines).


Another common observation on such sonotrodes is the presence of small
streamers on their lateral side, visible near the upper left corner of
Fig.~\ref{figcone}b.  As seen in the right part of the latter figure,
this phenomenon is reasonably caught by the simulation, and this is
the reason why the deformation of the transducer was accounted for.
Indeed, our result suggests that such small structures result from the
lateral vibration of the sonotrode, which emits a radial wave, and
produces a small zone of large acoustic pressure. The bubbles in this
zone strongly attenuate the wave, and produces a traveling part in the
radial wave.  The physical mechanism is therefore similar to the cone
formation, but here the stagnation point is very close to the
sonotrode surface, so that only a small flat filamentary structure is
formed.

\begin{figure}[h!t]
  \centering
  \includegraphics[width=0.9\linewidth]{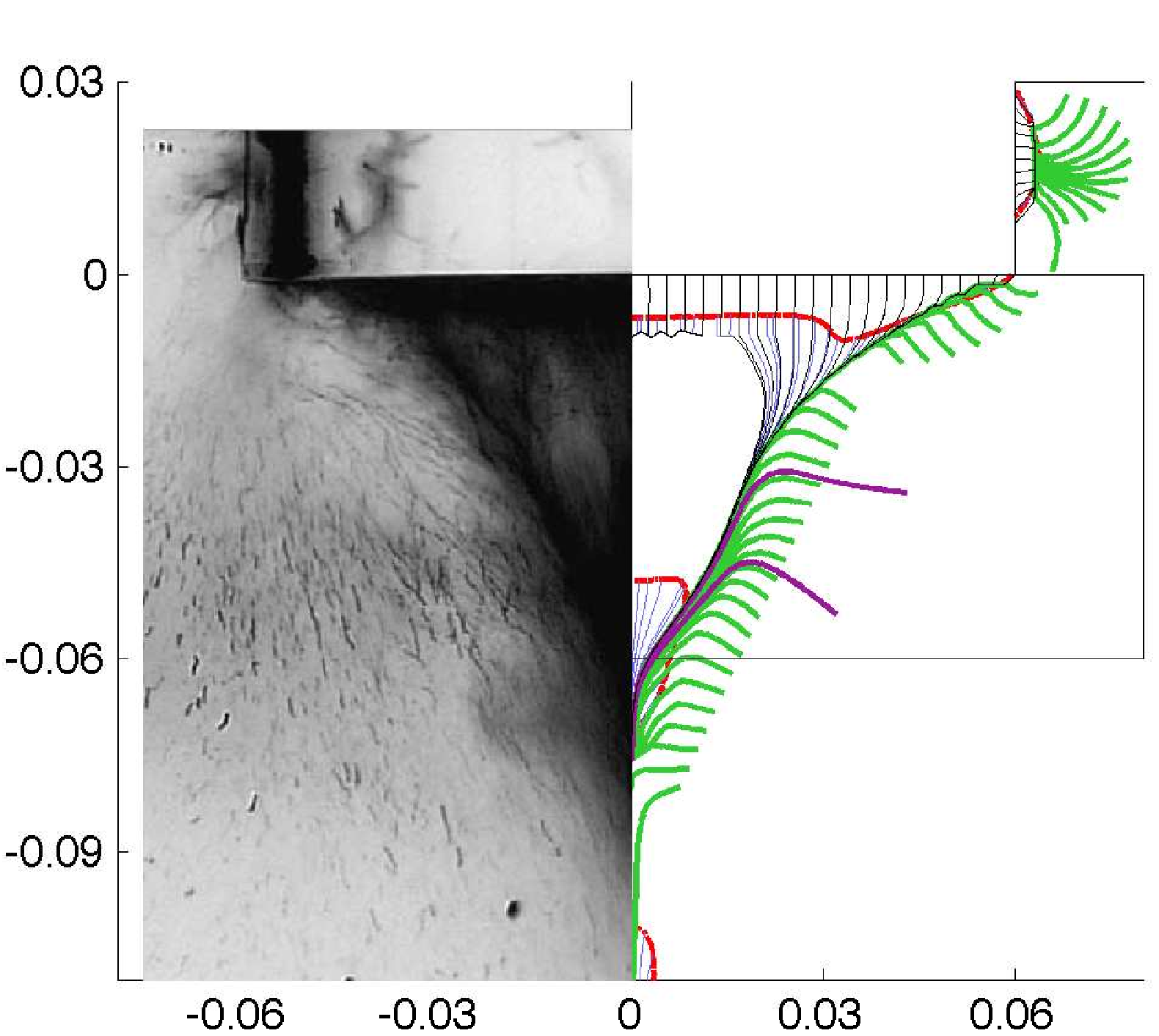}
  \caption{Same as Fig.~\ref{figcone}b, sketching additional streamers
    originating from arbitrary points points below the Blake threshold
    (green and magenta lines).}
  \label{figconeadd}
\end{figure}

Figure~\ref{figconeadd} shows the same result as Fig.~\ref{figcone}b
where we sketched additional streamers originating from arbitrary
points in the liquid (green lines), which makes the comparison of the
lateral filamentary structure with experiments more striking.
Furthermore, this representation allows to evidence streamers starting
near the cone lateral boundary and quickly merging with the latter, as
indeed visible on the experimental picture. Figure~\ref{figconeadd}
also shows that the corner of the sonotrode acts as a separatrix
between the streamers attracted by the cone and the ones attracted by
the lateral filamentary structure, as can be also speculated from the
experimental picture. We note however that streamers starting at a
larger distance from the cone (see magenta lines in
Fig.~\ref{figconeadd}) start upwards, whereas such streamers on the
experimental picture seems to start downwards. The experimental image
also suggests that some streamers starting from points far from the
cone (say, near $r=0.06$~m, $z=-0.07$~m) seem to be attracted by
points located outside the picture, and this feature is not caught by
our simulation.

\begin{figure}[h!t]
  \centering
  \includegraphics[width=\linewidth]{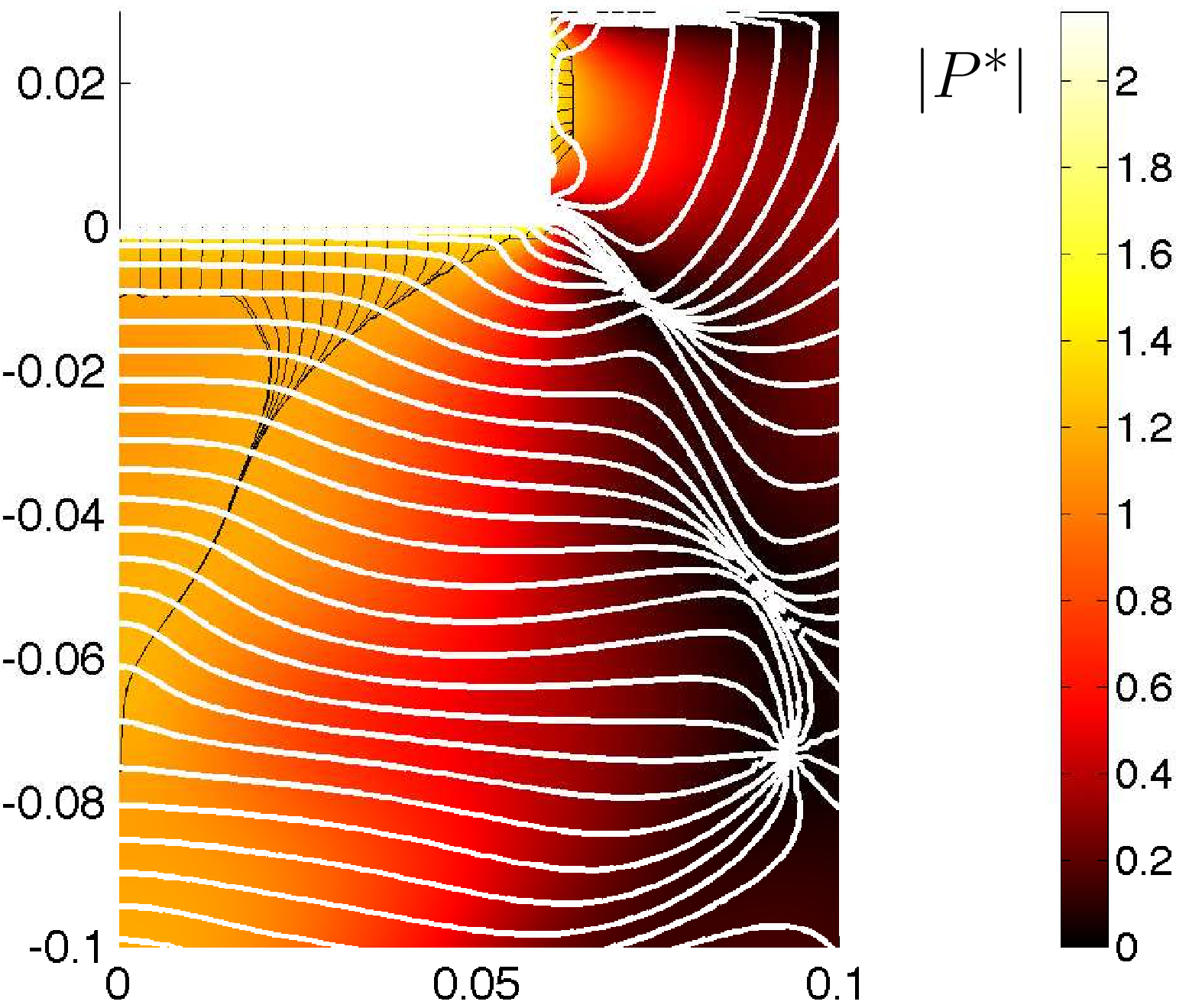}
  \caption{Color plot of the acoustic pressure field in the liquid.
    The white lines are the contour lines of the phase $\phi$ of the
    acoustic pressure field (see Eq.~(\ref{defpaco})). The
    {\sstreamers} are recalled by the black lines.}
  \label{figconesurf}
\end{figure}

The interpretation of the cone structure as the result of the
combination of a longitudinal traveling wave and a lateral standing
wave proposed in Ref.~\cite{kochmettin2004} is therefore confirmed by
the present model. However, the above analysis does not tell much
about how such an acoustic field appears. Since cone bubble structure
are very robust against amplitude, sonotrode size (see discussion in
Ref.~\cite{moussatovgrangerdubus2003}), and even appear near the walls
of ultrasonic baths (see Ref.~\cite{mettin2005}, and next section),
there must be some generic mechanism responsible of its formation.
Since traveling waves constitute the key phenomenon of the problem, it
is interesting to sketch the contour lines of the acoustic field phase
($\phi$ in Eq.~(\ref{defpaco})). For a traveling wave these lines are
orthogonal to the direction of propagation and constitute therefore a
powerful visual tool to assess the latter. The result is displayed in
Fig.~\ref{figconesurf} (white lines). It can be seen that, while the
wave mainly propagates along the $z$-direction inside the cone, it
bends into an oblique direction near the cone boundary, targeting at a
point located on the symmetry axis ({\sstreamers} are recalled in
black).
We emphasize that the emission of an oblique wave from the sonotrode
corner was found to occur in all our simulations, whatever the sizes
of the sonotrode and the liquid domain, and the type of the bottom
liquid boundary. More importantly, we found that the same phenomenon
occurs whenever the deformation of the sonotrode was accounted for or
not, which rules out any effect of a non uniform displacement of the
sonotrode tip.


\begin{figure}[h!t]
  \centering
  \includegraphics[width=\linewidth]{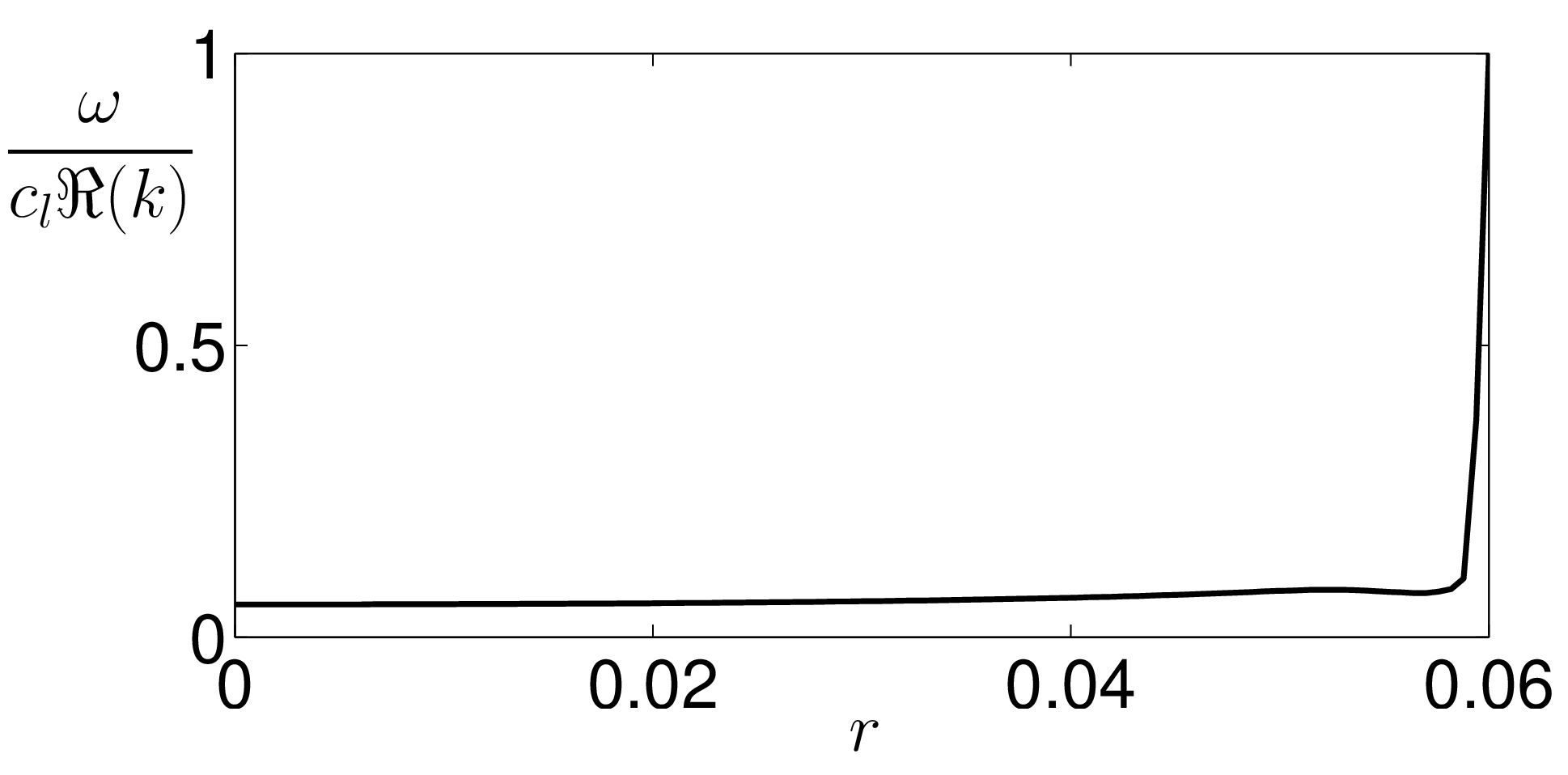}
  \caption{Radial profile of sound velocity on the transducer
    area. The sound velocity is non-dimensionalized by the sound
    velocity in the pure liquid $c_l$. }
  \label{figconeprofceff}
\end{figure}

We therefore infer that the slanting of the wave propagation direction
is definitely linked to the presence of strongly driven bubbles near
the vibrating area. These bubbles dissipate a lot of energy, rendering
the square of the local wave number almost purely imaginary. But the
acoustic field near the outer points of the sonotrode is weaker,
because this region is less constrained laterally (see pressure
profile in thick solid line in Fig.~\ref{figconeprofpa}). Therefore,
as evidenced in {\firstpaper} (see Fig.~4 in the latter reference),
the real part of $k$ is higher in the central part of the sonotrode
than in its outer part, and the opposite holds for the sound velocity
$\omega/\Re(k)$, which is confirmed by Fig.~\ref{figconeprofceff}.
There appears therefore an outward gradient of sound speed along the
sonotrode area, which, following Huygens principle, bends the wave
number towards the axis, and produces the conical traveling wave
visible on Fig.~\ref{figconesurf}. This traveling wave produces in
turn a strong Bjerknes force directed along the propagation direction
of the wave (a large $I_S$ term in Eq.~(\ref{defbjerknesintter})),
which structures the bubbles into a conical shape. The latter scenario
was qualitatively checked by simple linear acoustics simulations:
setting the sound field to $c_l$ uniformly in the liquid except in a
thin cylinder-shaped region below the transducer, the bending of the
iso-$\phi$ lines was indeed observed.

\olchange{A remarkable feature of cone bubble structures is the
  invariance of their shape when intensity is increased above a given
  level \cite{moussatovgrangerdubus2003}. Following the suggestion of
  an anonymous reviewer, we performed an additional simulation of the
  above configuration, increasing the sonotrode displacement $U_0$ up
  to 4.2~$\mu$m (instead of 1.4 $\mu$m).  The results are presented as
  supplementary material.  The cone shape obtained is indiscernible
  from the one presented in Fig.~\ref{figconeadd}.  However, a close
  examination of the axial pressure profiles in the two cases reveals
  that the acoustic fields differ mainly in a thin layer of about 5~mm
  near the sonotrode (see supplementary material). This is a clear
  manifestation of the self-saturation effect inherent to the present
  model through the field-dependence of the attenuation coefficient
  (see \firstpaper): increasing the sonotrode displacement produces a
  large increase of acoustic pressure only locally, but the bubbles in
  this zone being excited more strongly, they dissipate the excess
  acoustic energy very rapidly. It should be noted that experimental
  manifestations of this phenomenon have been reported in the early
  work of Rozenberg \cite{rozenbergchap71}.

Other studies demonstrates that conversely, for low excitations, the
shape of the cone bubble structure does depend on the driving.
Although some of our simulations could partially catch such a
dependence, convergence problems in this range prohibited any firm
conclusion. We observed however that the cone shape was very sensitive
to the choice of the bubble density when the latter was low enough.
This might suggest that the experimentally observed shape dependence
for low drivings would be rather due to the variation of the bubble
density with acoustic pressure, than to the pressure dependence of a
single bubble dissipation. This is a missing brick in our model since
we consider constant bubble densities above the Blake threshold. We
thus infer that the model in its present form is not able to catch
the latter experimental feature. 

One should finally mention, as underlined in
Ref.~\cite{dubusvanhille2010}, another explanation for the cone
structure robustness against driving level above a certain threshold,
borrowed to phase transitions theory. Skokov and co-workers measured
laser intensity transmission through a cavitation zone and obtained
time-series presenting fluctuations whose power spectrum were found to
be inversely proportional to frequency \cite{skokov2006,skokov2007}.
This feature, also termed as ``flicker noise'' has been shown to occur
in various physical processes \cite{bak88}, and has been interpreted
as a consequence of the interaction between two phase transitions, one
subcritical, the other supercritical \cite{skokov2001}. The result is
that the system self-organizes into a critical state, whatever the
precise value of the controlling parameter, contrarily to classical
critical states which require a fine tuning of the latter to be
reached.  This phenomenon has been termed as ``self-criticality'' and
produces organized self-similar spatial structures, reminiscent of
bubble web-like organization.  Self-criticality can be modeled
generically by a stochastic dynamical system of two equations,
generalizing the Ginzburg-Landau equation.  One of the two order
parameters exhibits fluctuations between two attractors, with a power
spectrum varying in~$1/f$, as observed in experiments. This original
theory has the advantage to explain some generally overlooked features
of cavitation clouds by a universal physical mechanism. However it
still remains very far from the precise cavitation physics, and the
proposed equations are phenomenological.  In particular the precise
physical sense of the order parameters remains to be explicited, maybe
on the light of the coupled evolutions of the bubble field and the
acoustic wave. For now, the model is still far from a predictive tool
for acoustic cavitation, but this promising approach remains opened.}

\olchange{\subsubsection{Experiments with a 7~cm diameter sonotrode}}
\olchange{Reference \cite{moussatovgrangerdubus2003} also present
  results for thinner sonotrodes, but the latters produce acoustic
  currents which deform the cone structure, so that a direct
  comparison of the experimental images and simulated cone structures
  is not possible in this case. In spite of the latter restriction, we
  performed additional simulations for a 8~mm diameter sonotrode
  (referred as ``type B'' in Ref.~\cite{moussatovgrangerdubus2003}),
  in an otherwise identical geometry, for 2 $\mu$m bubble radii and a
  bubble density $N_0=$~90 bubbles/mm$^3$, in order to match at best
  the experimental shape of the cone structure. The result of the
  visual comparison between the simulated cone structure and Fig.~2 in
  \cite{moussatovgrangerdubus2003} is deferred to supplementary
  material, and shows that, in spite of the blurring of the structure
  by acoustic currents, some similarities in the cone shape can be
  observed.

  Apart from imaging cone structures, Dubus and co-workers have
  collected valuable quantitative experimental informations in
  additional studies, using 7~cm diameter sonotrodes excited in pulsed
  mode in order to avoid acoustic currents
  \cite{camposdubus2005,dubusvanhille2010}. In order to put our model
  to the test, we performed additional simulations of a 7~cm
  sonotrode, again with for 2 $\mu$m bubble radii and $N_0=$~90
  bubbles/mm$^3$.  The transducer displacement was set in order to
  match the velocity of the sonotrode area measured in
  Ref.~\cite{camposdubus2005} (1.31 m/s).  It is interesting to
  compare the calculated and experimental axial acoustic pressure
  profiles (Fig~6 in Ref.~\cite{camposdubus2005}, star signs). The
  result is displayed in Fig.~\ref{figpazcone7cm}. It can be seen that
  the agreement is somewhat poor: even if the order of magnitude of
  the predicted pressure field is reasonable, it vanishes much more
  slowly than the experimental one. We increased the bubble density up
  to $N_0=$~360 bubbles/mm$^3$ without appreciable change in the
  pressure profile, except near the sonotrode. This underlines the
  limit of the present model, and may result of the rough assumption
  of a constant bubble density.  Besides, the pressure field
  calculated in the immediate vicinity of the transducer ($z=0$) is
  found to be much larger than the experimental one. It should however
  be noted that hydrophones have a finite size, which does not allow
  to measure the acoustic pressure exactly near $z=0$, where the
  present theory predicts a large pressure gradient.

    \begin{figure}[Htb]
      \centering
      \includegraphics[width=\linewidth]{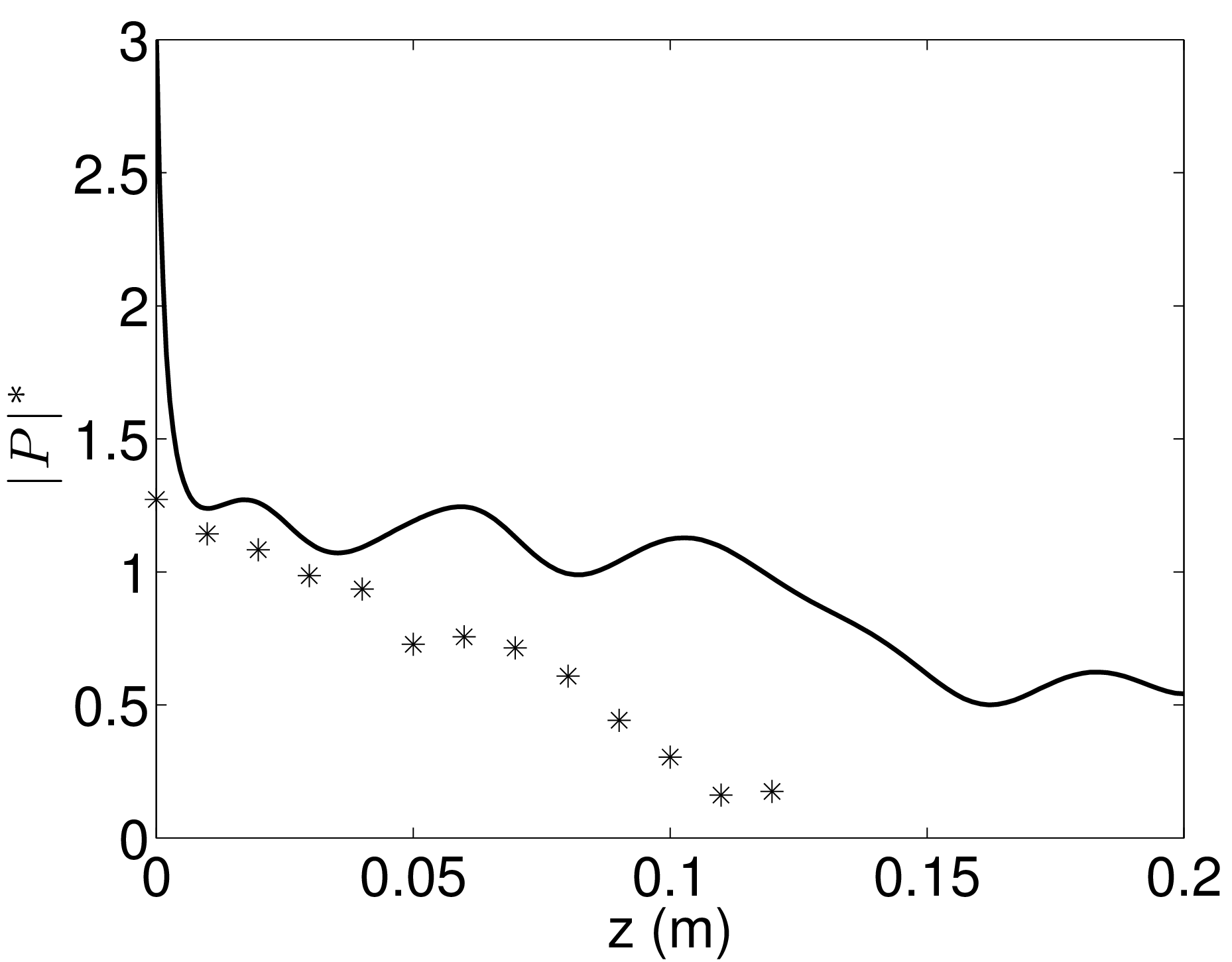}
      \caption{Pressure profile on the symmetry axis for a 7 cm
        diameter sonotrode, with $R_0=2$~$\mu$m,
        $N_0=90$~bubbles/mm$^3$. The velocity of the sonotrode tip is
        1.31 m/s. Solid line: prediction by the present model. Stars:
        experimental measurements redrawn from Fig.~6 in
        Ref.~\cite{camposdubus2005}.}
      \label{figpazcone7cm}
    \end{figure}
}

\olchange{Finally, Dubus and co-workers \cite{dubusvanhille2010}
  proposed an alternative explanation of the bubble cone structure.}
Their analysis relies on the formation of a nonlinear resonant layer
of bubbly liquid attached on the transducer, the focusing being
qualitatively attributed to the curvature of the bubble layer.
\olchange{This layer produces a phase shift of the wave emitted by the
  sonotrode, dependent on its local width. The argumentation is
  supported by phase measurements of the pressure field in a
  horizontal plane below the layer.  We have therefore reported the
  corresponding experimental values, along with the prediction of our
  simulation in Fig.~\ref{figtaudubus} (the simulation parameters are
  the same as for Fig.~\ref{figpazcone7cm}). It can be seen that the
  agreement is very good, and the result was found almost unsensitive
  to an increase of the bubble density up to $N_0=$~360 bubbles/mm$^3$.

    \begin{figure}[Htb]
     \centering
     \includegraphics[width=\linewidth]{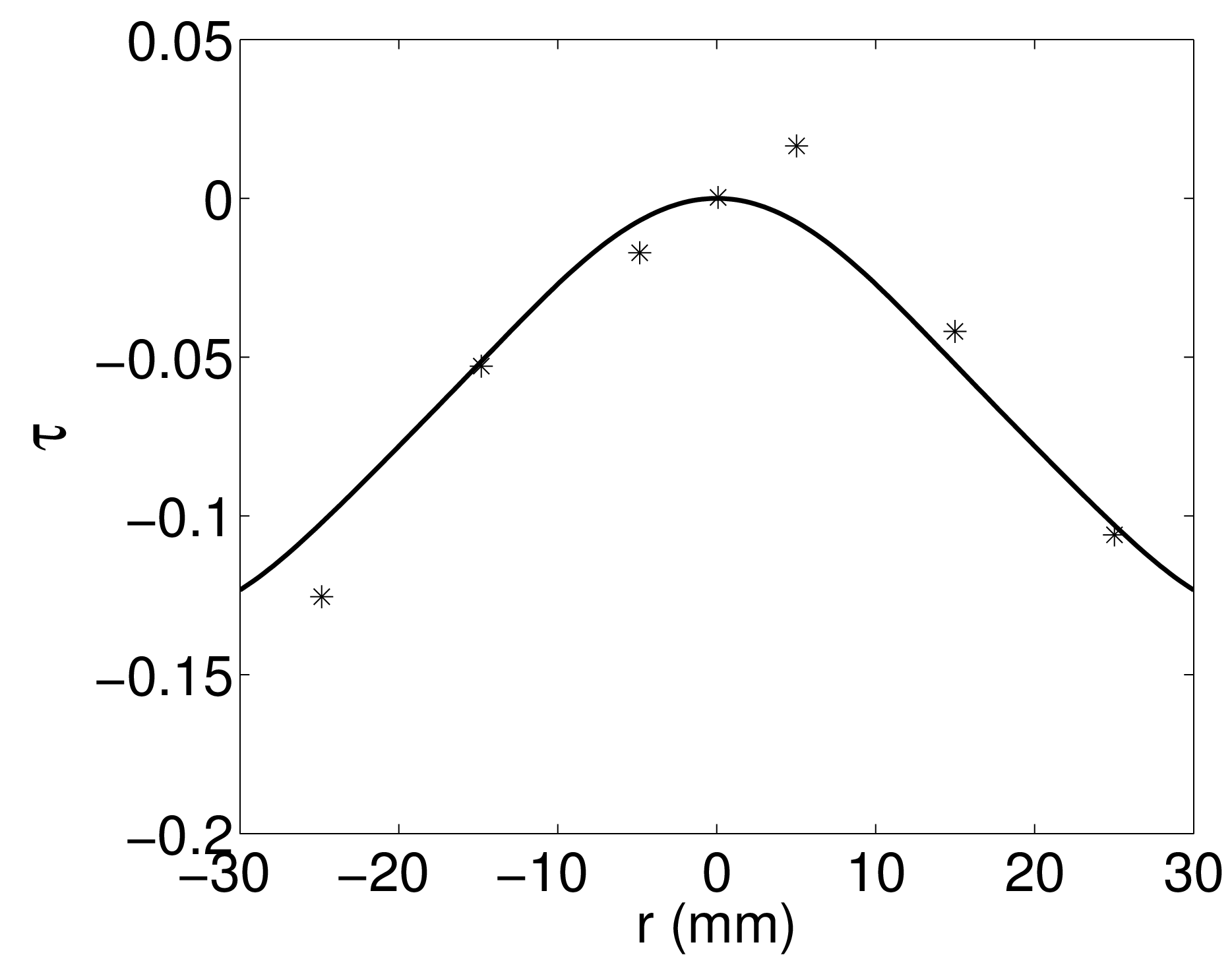}
     \caption{Radial variation of the phase of the pressure field, at
       1.5~cm from the transducer. The quantity represented is
       $(\phi(0,1.5\text{ cm})-\phi(r,1.5\text{ cm}))/(2\pi)$, where
       $\phi(r,z)=\arg(p(r,z))$. The solid line represents the
       theoretical prediction, and the stars are experimental results
       redrawn from Fig.~10 in Ref.~\cite{dubusvanhille2010}. The
       simulations parameter are the same as for
       Fig.~\ref{figpazcone7cm}.}
     \label{figtaudubus}
   \end{figure}
}

\olchange{The present calculation shows therefore that the existence
  of a curved resonant layer of bubble is not necessary to explain the
  experimental results, although nonlinear resonant effects cannot be
  discarded. The} wave focuses anyway just because the outermost
bubbles in the bubble layer are more smoothly driven than the central
ones, and therefore have a lower influence on the local sound speed.
This a purely nonlinear effect, and is correctly summarized by Fig.~4
in {\firstpaper}. Besides, our simulations indeed exhibit a bubbly
layer near the transducer, but rather than being resonant, this layer
is in fact found to be very dissipative. Such a large dissipation
cannot be predicted by using a reduced bubble dynamic equation, and is
strongly correlated to the inertial character of the bubble
oscillations, as shown in our companion paper {\firstpaper}.

\subsection{Cleaning bath}
\begin{figure}[ht]
  \centering
  \includegraphics[width=\linewidth]{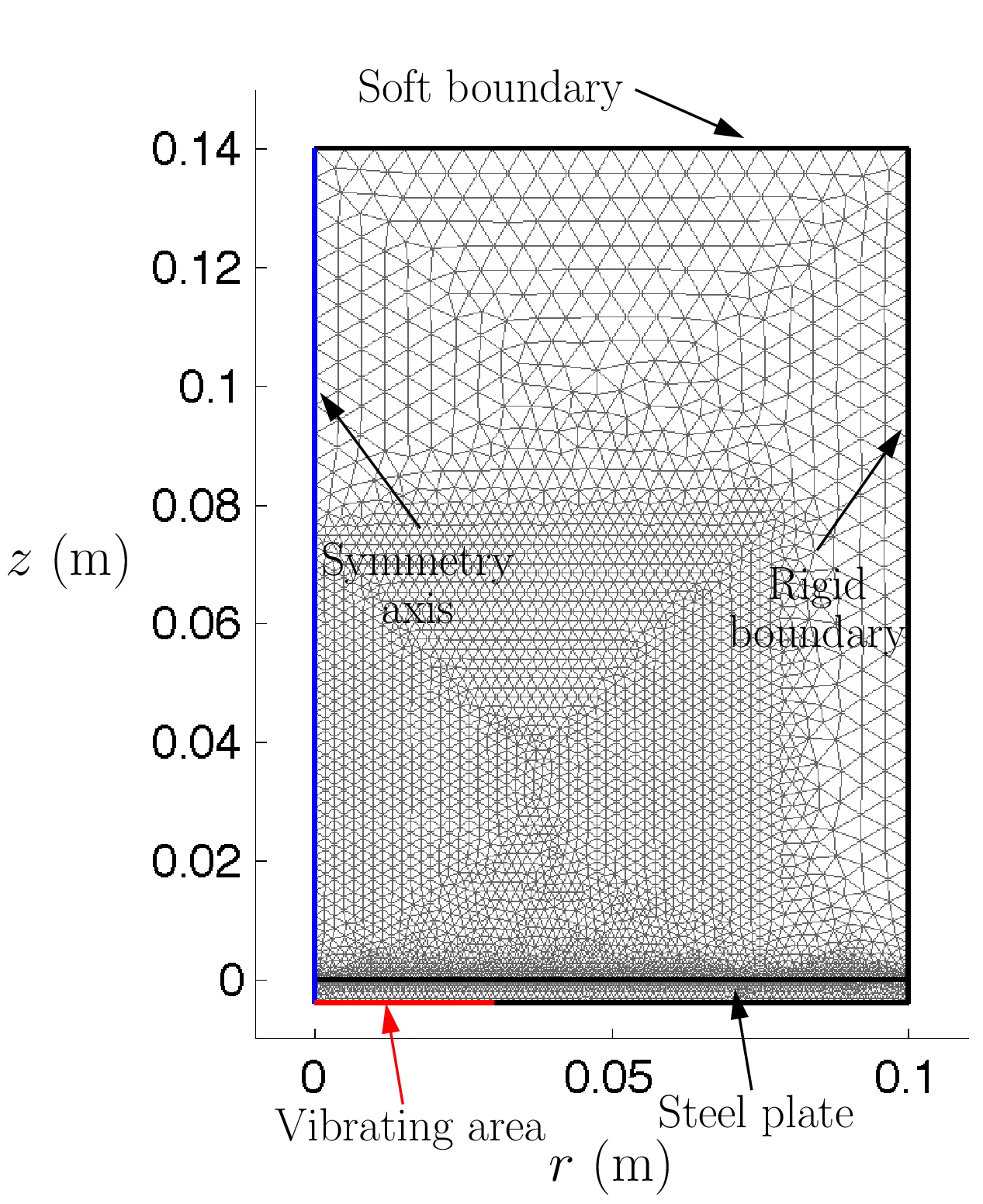}
  \caption{Geometry and meshing of the cleaning bath configuration.
    The geometry has axial symmetry and only a half-plane cut is
    represented. The blue line is the symmetry axis. A uniform
    displacement $U_0$ is imposed on the red line.}
 \label{figbathmesh}  
\end{figure}

The system considered is axi-symmetrical and is represented in a
half-plane cut on Fig.~\ref{figbathmesh}. The bottom of the bath is a
thin circular steel plate of 4 mm thickness and 20~cm diameter, which
upper side is in contact with the liquid, while its lower side is
free, except its central part (red line on Fig.~\ref{figbathmesh},
$r\in [0, 3$~cm$]$) which is assumed to vibrate with a uniform
amplitude $U_0 =2$~$\mu$m and a frequency of 20~kHz. This boundary
condition is a simplification of a system where a piezo-ceramic ring
would be clamped to the bottom side of the plate and impose an
oscillatory displacement. The liquid fills the space above the plate,
and is limited laterally by a cylindrical boundary, which is assumed
infinitely rigid, while the free surface is considered as infinitely
soft. The plate is assumed elastic and its deformation follows the
Hooke's law, its vibrations being coupled to the acoustic field in the
liquid by adequate interface conditions \cite{louisnardgonzalez2009}.

\begin{figure}[h!t]
  \centering
  \includegraphics[width=\linewidth]{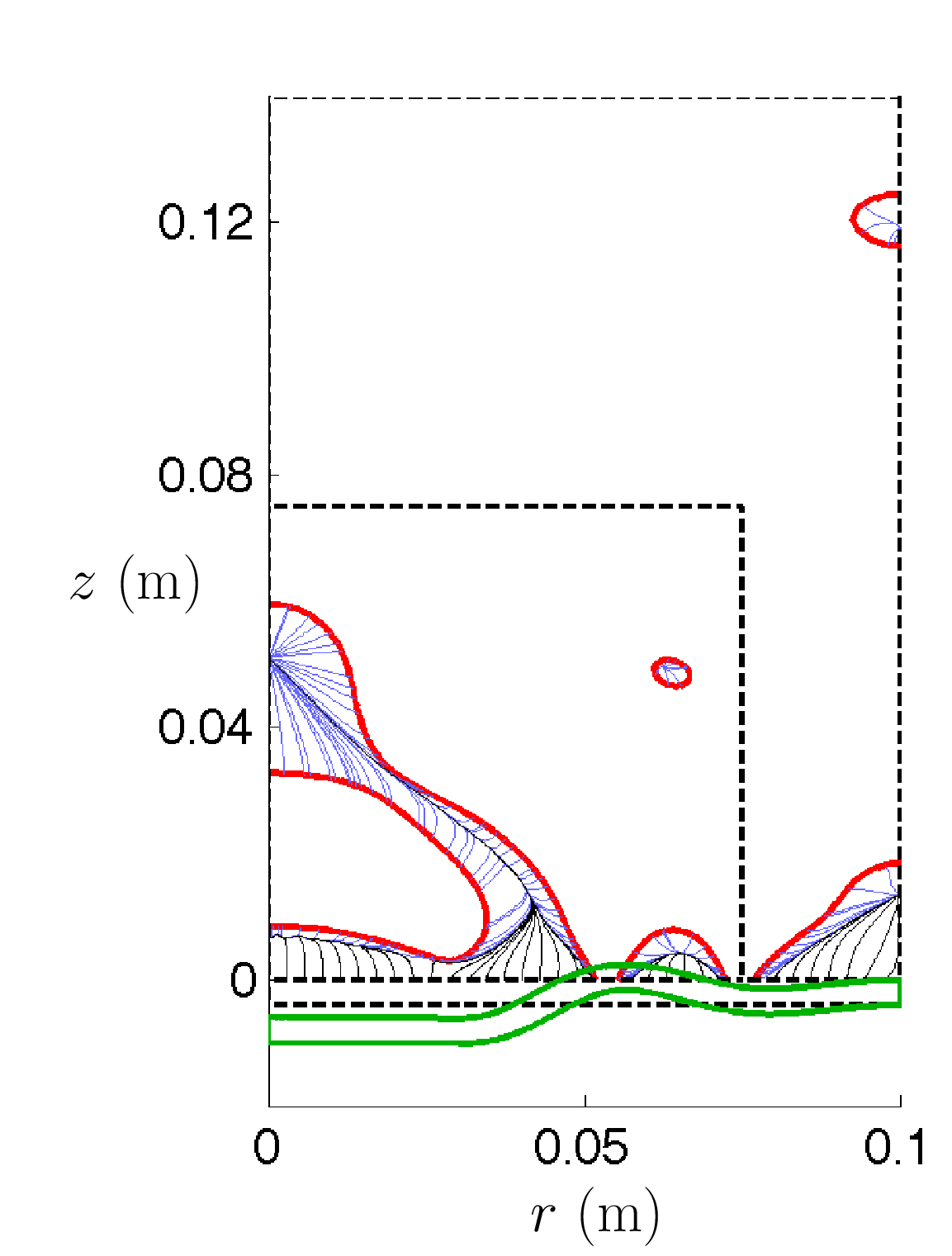}
  \caption{Simulation of the cleaning bath configuration. The thick
    red lines materialize the Blake-threshold. The black lines are the
    {\sstreamers}, paths of bubbles originating from the vibrating
    plate, and the blue lines are the {\lstreamers}, path of bubbles
    originating from the Blake threshold. The green thick line
    represents the deformed shape of the plate at $\omega t=2n\pi$.}
 \label{figbathmain}  
\end{figure}

Fig.~\ref{figbathmain} displays the result obtained with 5~$\mu$m air
bubbles and a water height of 0.14~m. The green line represents the
deformed shape of the bottom plate, and shows that a flexural standing
wave is excited in the latter. This flexural wave produces a spatially
inhomogeneous acoustic pressure field on the plate area, ranging in
this case between 0.5 and 2.8 bar (the red lines represent the Blake
threshold contour curves). In the zones on the plate where the
acoustic pressure is larger than the Blake threshold, bubbles are
produced, dissipating locally a large energy and modifying the sound
speed. The remaining mechanisms are similar to the cone bubble
structure described above. Dissipation produces traveling waves, and
and can even result into the production of a small cone bubble
structure (near $r=0.04$~m) attached on the plate, while other
structures on the plate are more similar to the above-mentioned
streamers attached on the lateral side of sonotrodes. A few streamers
are also visible in the middle of the liquid, located on the antinodes
of the wave, which again takes a standing character in this region.

\begin{figure}[ht]
  \centering
  \includegraphics[width=\linewidth]{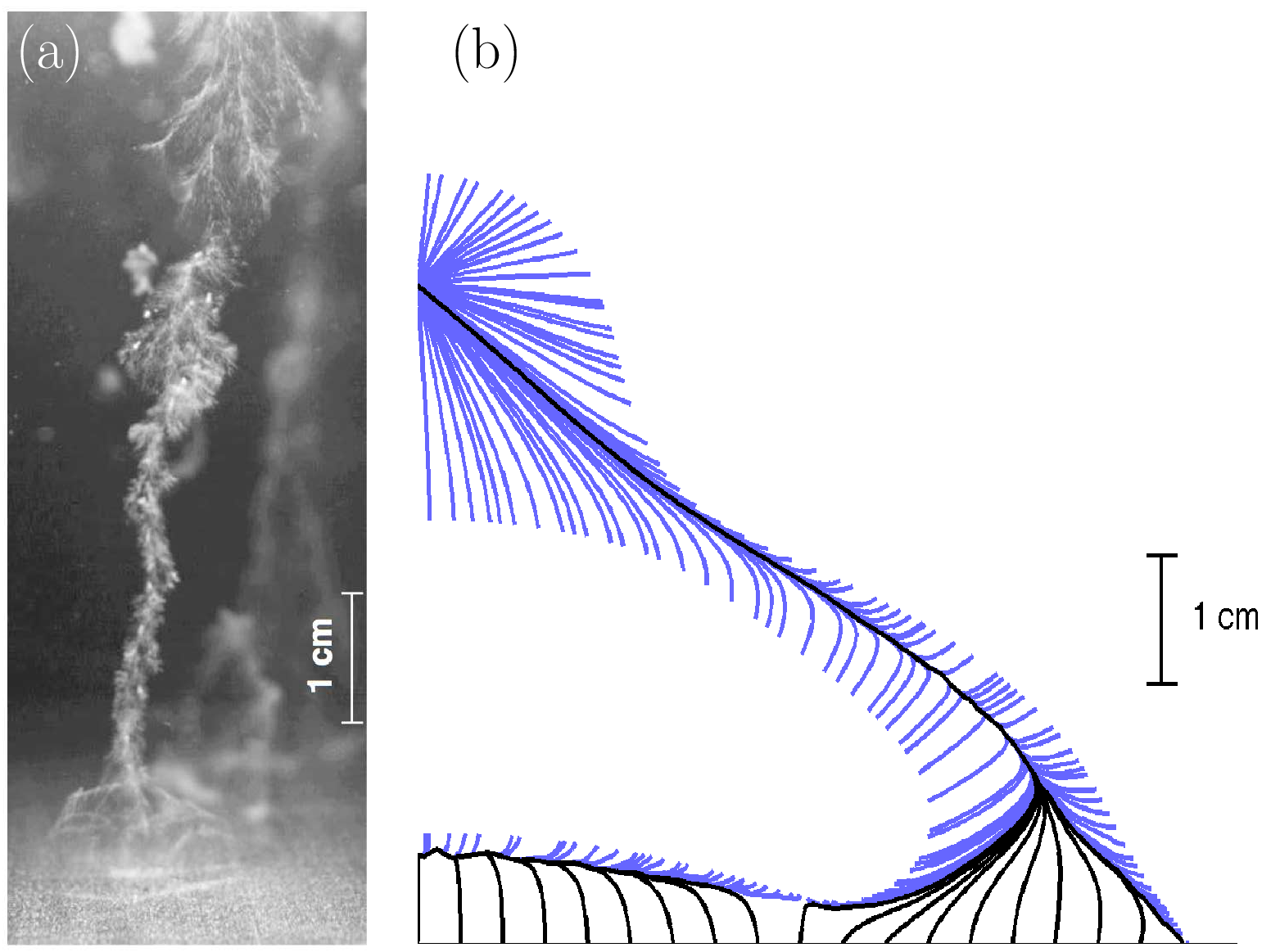}
  \caption{Comparison between: (a) a flare structure observed on the
    lateral side of a cleaning bath (reprinted from
    Ref.~\cite{mettin2005} with permission of Research Signpost; the
    original image has been 90$^\circ$ rotated for comparison
    purposes) (b) the structure predicted in the simulated
    configuration (zoom of Fig.~\ref{figbathmain} with more
    {\lstreamers} sketched).}
 \label{figflare}  
\end{figure}

An interesting feature can also be seen in Fig.~\ref{figbathmain} and
is magnified in Fig.~\ref{figflare}b. It is seen that the
{\sstreamers} (black lines) merging at the vertex of the cone then
follow a unique path up to a stagnation point located on the symmetry
axis. If we also launch bubbles from the Bjerknes contour curves, the
corresponding {\lstreamers} (blue lines) join the main {\sstreamers}
up to the stagnation point where numerous secondary streamers end up,
forming a star-like structure. The experimental occurrence of this
behavior has been reported in a detailed presentation by Mettin
\cite{mettin2005}, who termed this bubbles arrangement as ``flare
structure'', and is represented in Fig.~\ref{figflare}a). It is seen
that, apart from the curvature of the structure observed in our
simulation, the latter reproduces reasonably well the main features of
the phenomenon, especially the cone structure attached to the
vibrating plate merging into a jet, fed laterally by bubbles
originating in the bulk liquid.  Mettin reported that this structure
was universally found in cleaning bath setups and proposed a
qualitative explanation involving a ``complicated near field structure
with shares of both traveling and standing waves''. The present
results seem to enforce this interpretation: a traveling wave takes
birth near the vibrating area, because of the strong attenuation by
the bubbles located there, and launches the bubbles far from the
plate. It ends up at a pressure antinode which attracts all the
bubbles, either coming from the plate, or taking birth in neighboring
zones. The lateral enrichment of the main bubble path originates from a
standing wave in the direction perpendicular to the traveling
wave.

\section{Summary and discussion}
\label{secconclusion} 

The model proposed in the companion paper {\firstpaper} has been
applied to classical 2D configurations, namely large area transducer
emitting in a liquid, and cleaning baths.  The density of bubbles was
assumed constant in zones where the acoustic pressure is above the
Blake threshold, and null everywhere else. The possible paths of the
bubbles were assumed to originate either from the vibrating parts of
the solid, or from the Blake threshold contour curves, and calculated
by computing the primary Bjerknes force field directly from nonlinear
bubble dynamics simulations.

In the case of large area transducer, cone bubble structures observed
experimentally can be easily reproduced, with reasonable agreement in
the cone shape. A strongly dissipative bubble layer was found to appear
near the transducer, and the cone boundaries were shown to be made of
bubbles following a focused traveling wave. The focusing was found to
result from a radial acoustic pressure gradient on the transducer
area, which, following the result of the companion paper
{\firstpaper} produces a radial gradient of sound velocity.
The streamers located on the lateral boundary of the transducer,
observed experimentally, were correctly reproduced by our simulations,
provided that the elastic deformations of the emitting transducer were
accounted for. Besides, the calculated stagnation point on the symmetry axis
was found to be much farther than the cone tip, which would explain
the long bubble tail visible in experimental picture.  

In the cleaning bath configurations, the flexural vibrations of the
bottom plate were found to produce several zones of large acoustic
pressures, located near the plate displacement antinodes.  These high
acoustic pressures produce locally a thin layer of strongly
oscillating bubbles dissipating a lot of acoustic energy, which yields
damped traveling waves and cone-like structures. In some cases, the
bubbles reaching the cone tip carry on their motion along a unique
line, ending into a distant pressure antinode, and laterally enriched
by bubbles originating from liquid zones excited above the Blake
threshold. The obtained structure is reminiscent of a flare-like
structure described in the literature, and known to occur frequently
in cleaning baths configurations.

The reasonable success of our model in predicting ab initio such
structures is encouraging, and seem to show that strong energy
dissipation by inertial bubbles is a key mechanism ruling the
structure of the acoustic field in a cavitating medium.  \olchange{It
  is interesting to note that the self-action of the acoustic field
  evidenced in the present paper differs from the mechanisms presented
  in the work of Kobelev \& Ostrovski \cite{kobelevostro89}. In the
  latter work, the self-action of the acoustic field is mediated by
  its slow influence on the bubble population, while here, the
  mechanism is only due to the bubbles radial motion, even for
  constant bubble density. The latter assumption constitutes however a
  weak point of our model, and requires }the arbitrary choice of two
free parameters: the ambient radius of the bubbles $R_0$, and the
bubble density, both being assumed spatially homogeneous in regions
above the Blake threshold. A more realistic model would require at
least the spatial redistribution of the bubbles. This may be done for
example by coupling the nonlinear Helmholtz equation used in this
paper with a convection-like equation for the bubble number density,
\olchange{as done in the linear case in Ref.~\cite{kobelevostro89},
  which requires the correct estimation of the average translational
  velocity of inertial bubbles. As already mentioned in introduction,
  this translational motion is described by a somewhat elaborate
  physics \cite{doinikov2005review,mettindoinikov2009}, and the
  estimation of an average velocity raises the} complicated issue of
properly averaging the translation equation
\cite{reddyszeri2002bjerknes,kreftingtoilliez2006,toilliezszeri2008}.
\olchange{Along the same line of investigation, an extension of the
  present work could consist in launching some bubbles in the acoustic
  fields presented here, and to calculate their paths within the
  latter by integrating in time the coupled equations of radial and
  translational motion described in
  Refs.~\cite{doinikov2005,mettindoinikov2009}. Apart from testing the
  validity of the current approximation, this would also provide a
  clear picture of the dynamics of bubble drift in the studied
  structures. This may reveal some unexpected features, such as
  bubble precession around some points in the liquid, as evidenced in
  simple standing waves in Ref.~\cite{mettindoinikov2009}.}  Such
refinements may be the subject of future work, keeping however as a
main objective a model simple enough to be used in real engineering
applications.
  
\olchange{Finally, many other bubble structures can be observed
  experimentally \cite{mettin2005}. Among the latter, the grouping of
  bubbles into so-called ``clusters'' classically appear in numerous
  experimental configuration either in the bulk liquid, or as
  hemi-spherical structures near solid boundaries, especially in the
  case of focused ultrasound \cite{hallez2010,chen2007,brujan2011}. In
  some aspects, they behave as a single large bubble and may collapse
  as a whole, emitting a complex set of primary and secondary
  shock-waves (see Ref.~\cite{brujan2011} for high-speed photographs),
  and yield strongly erosive effects \cite{krefting2004}.  A
  complete theoretical description of such structures is not yet
  available, the difficulty lying in the strong interaction between
  the bubbles constituting the cloud. Clearly, this interaction cannot
  be accounted for by our simple model which includes only primary
  Bjerknes forces. Moreover, these structures often present a
  transitory behavior, moving as a whole entity in the liquid,
  appearing and disappearing near other structures
  \cite{krefting2004,mettin2005}, and have even been observed as early
  precursors of conical structures \cite{moussatovmettin2003}. As a
  conclusion, we feel therefore that our model in the present
  form cannot account for such structures.}

\section{Acknowledgments}
The author acknowledges the support of the French Agence Nationale de
la Recherche (ANR), under grant SONONUCLICE (ANR-09-BLAN-0040-02)
''Contr\^ole par ultrasons de la nucl\'eation de la glace pour
l'optimisation des proc\'ed\'es de cong\'elation et de
lyophilisation''. Besides the author would like to thank Nicolas Huc
of COMSOL France for his help in solving convergence issues.

\appendix
\olchange{\section{Supplementary data}
Supplementary data associated with this article can be found, in the
online version, at doi:xx.xxxx/x.xxx.xxxx.xx.xx.}

\vfill
\pagebreak
 
\bibliographystyle{elsart-num-usson}



\pagebreak
\listoffigures

\end{document}